\documentclass[10pt,letterpaper]{article}
\usepackage[top=0.85in,left=2.75in,footskip=0.75in]{geometry}

\usepackage{amsmath,amssymb}

\usepackage{changepage}

\usepackage[utf8x]{inputenc}

\usepackage{textcomp,marvosym}

\usepackage{cite}

\usepackage{nameref,hyperref}

\usepackage[right]{lineno}

\usepackage{microtype}
\DisableLigatures[f]{encoding = *, family = * }

\usepackage[table]{xcolor}

\usepackage{array}

\newcolumntype{+}{!{\vrule width 2pt}}

\newlength\savedwidth

\newcommand\thickhline{\noalign{\global\savedwidth\arrayrulewidth\global\arrayrulewidth 2pt}%
\hline
\noalign{\global\arrayrulewidth\savedwidth}}

\raggedright
\usepackage[aboveskip=1pt,labelfont=bf,labelsep=period,justification=raggedright,singlelinecheck=off]{caption}

\makeatletter
\renewcommand{\@biblabel}[1]{\quad#1.}
\makeatother

\usepackage{lastpage,fancyhdr,graphicx}
\usepackage{epstopdf}
\fancyheadoffset[L]{2.25in}
\fancyfootoffset[L]{2.25in}
\begin{document}
\vspace*{0.2in}

\begin{flushleft}
{\Large
\textbf\newline{Contrasting the effects of adaptation and synaptic filtering on the timescales of dynamics in recurrent networks} 
}
\newline
\\
Manuel Beiran\textsuperscript{1*},
Srdjan Ostojic\textsuperscript{1*}
\\
\bigskip
\textbf{1} Group for Neural Theory, Laboratoire de Neurosciences Cognitives Computationnelles, D\'{e}partement d'\'{E}tudes Cognitives, \'{E}cole Normale Sup\'{e}rieure, INSERM U960, PSL University, Paris, France
\\
\bigskip

%
%





* manuel.beiran@ens.fr (MB); srdjan.ostojic@ens.fr (SO)

\end{flushleft}
\section*{Abstract}
Neural activity in awake behaving animals exhibits a vast range of timescales that can be several fold larger than the membrane time constant of individual neurons. Two types of mechanisms have been proposed to explain this conundrum. One possibility is that large timescales are generated by a network mechanism based on positive feedback, but this hypothesis requires fine-tuning of the strength or structure of the synaptic connections. A second possibility is that large timescales in the neural dynamics are inherited from large timescales of underlying biophysical processes, two prominent candidates being intrinsic adaptive ionic currents and synaptic transmission. How the timescales of adaptation or synaptic transmission influence the timescale of the network dynamics has however not been fully explored.
To address this question, here we analyze large networks of randomly connected excitatory and inhibitory units with additional degrees of freedom that correspond to adaptation or synaptic filtering. We determine the fixed points of the systems, their stability to perturbations and the corresponding dynamical timescales. Furthermore, we apply dynamical mean field theory to study the temporal statistics of the activity in the fluctuating regime, and examine how the adaptation and synaptic timescales transfer from individual units to the whole population.
Our overarching finding is that synaptic filtering and adaptation in single neurons have very different effects at the network level. Unexpectedly, the macroscopic network dynamics do not inherit the large timescale present in adaptive currents. In contrast, the timescales of network activity increase proportionally to the time constant of the synaptic filter. Altogether, our study demonstrates that the timescales of different biophysical processes have different effects on the network level, so that the slow processes within individual neurons do not necessarily induce slow activity in large recurrent neural networks.

 %


\section*{Introduction}

	
	Adaptive behaviour requires processing information over a vast span of timescales \cite{Fairhall2001},  ranging from micro-seconds for acoustic localisation \cite{Grothe2010}, milliseconds for detecting changes in the visual field  \cite{Tchumatchenko}, seconds for evidence integration \cite{Smith2004} and working memory \cite{Miyashita1988}, to hours, days or years in the case of long-term memory. Neural activity in the brain is matched to the  computational requirements imposed by behaviour, and consequently displays dynamics over a similarly vast range of timescales \cite{Bair2004, Bernacchia2011, Murray2014}. Since the membrane time constant of an isolated neuron is of the order of tens of milliseconds, the origin of the long timescales observed in the neural activity has been an outstanding puzzle.
	
	Two broad classes of mechanisms have been proposed to account for the existence of long timescales in the neural activity. The first class relies on non-linear collective dynamics that emerge from synaptic interactions between neurons in the local network. Such mechanisms have been proposed to model a variety of phenomena that include working memory \cite{Wang2001}, decision-making \cite{Wang2008} and slow variability in the cortex \cite{Litwin-Kumar2012}. In those models, long timescales emerge close to bifurcations between  different types of dynamical states, and therefore typically  rely on  the fine tuning of some parameter \cite{Huang2017}. An alternative class of mechanisms posits that long timescales are directly inherited from long time constants that exist within individual neurons, at the level of hidden internal states \cite{Buonomano2009}. Indeed biophysical processes at the cellular and synaptic level display a rich repertoire of timescales. These include short-term plasticity that functions at the range of hundreds of milliseconds \cite{Zucker2002, Markram1998}, a variety of synaptic channels with timescales from tens to hundreds of milliseconds \cite{Newberry, Batchelor1994, Garthwaite1991,Lester1990}, ion channel kinetics implementing adaptive phenomena \cite{Johnston1995}, calcium dynamics \cite{Berridge2003} or shifts in ionic reversal potentials \cite{Gal2010}. How the timescales of these internal processes affect the timescales of activity at the network level has however not been fully explored.
	
	In this study, we focus on adaptative ion-channel currents, which are known to exhibit timescales over several orders of magnitude \cite{LaCamera2006, Benda2003, Ermentrout2001}. We contrast their effects on recurrent network dynamics with the effect of the temporal filtering of inputs through synaptic currents, which also expands over a large range of timescales \cite{Hennig2013}. To this end, we extend classical rate models \cite{Wilson1972, Wilson1973, Sompolinsky1988, Abbott1994} of randomly connected recurrent networks by including for each individual unit a hidden variable that corresponds to either the adapting of the synaptic current. We systematically determine the types of collective activity that emerge in such networks. We then compare the timescales on the level of individual units with the activity within the network.
	
\section*{Results}

We consider $N$ coupled inhibitory and excitatory units whose dynamics are given by two variables: the input current $x_i$ and a slow variable $s_i$ or $w_i$ that accounts for the synaptic filtering or adaptation current respectively. The instantaneous firing rate of each neuron is obtained by applying a static non-linearity $\phi\left(x\right)$ to the input current at every point in time. For simplicity, we use a positive and bounded threshold-linear transfer function
  
\begin{equation}
\phi\left(x\right) = \begin{cases}
\left[x-\gamma\right]^+ &\text{if } x-\gamma <\phi_{\text{max}} \\
\phi_{\text{max}}  &\text{otherwise},
\end{cases} \label{transfer}
\end{equation}

\noindent where $[\cdot]^+$ indicates the positive part, $\gamma$ is the activation threshold and $\phi_{\text{max}}$ the maximum firing rate.

 Single neuron adaptation is described by the variable $w\left(t\right)$ that low-pass filters the linearized firing rate  with a timescale $\tau_w$, slower than the membrane time constant $\tau_m$, and feeds it back with opposite sign into the input current dynamics (see \textit{Methods}). The dynamics of the $i$-th adaptive neuron are given by

\begin{equation}
\begin{cases}
&\tau_m \, \dot{x}_i \left(t\right)=-x_i\left(t\right)+\sum_{j=1}^N J_{ij}\phi\left(x_j\left(t\right)\right) -g_w w_i\left(t\right)+I_i\left(t\right)\\
& \tau_w \dot{w}_i\left(t\right) = -w_i\left(t\right) + x_i\left(t\right) - \gamma,
\end{cases} \label{sys_adap1}
\end{equation}

\noindent where $I_i\left(t\right)$ is the external input current to neuron $i$.
 
Synaptic filtering consists in low-pass filtering the synaptic input received by a cell with time constant $\tau_s$, before it contributes to the input current. The dynamics of the $i$-th neuron in a network with synaptic filtering are

\begin{equation}
\begin{cases}
&\tau_m \, \dot{x}_i \left(t\right)=-x_i\left(t\right)+s_i\left(t\right)\\
& \tau_s \dot{s}_i\left(t\right) = -s_i\left(t\right) + \sum_{j=1}^N J_{ij}\phi\left(x_j\left(t\right)\right)+I_i\left(t\right).
\end{cases} \label{sys_adap2}
\end{equation}

The matrix element $J_{ij}$ corresponds to the synaptic coupling strength from neuron $j$ onto neuron $j$. In this study we focus on neuronal populations of inhibitory and excitatory units, whose connectivity is sparse, random, with constant in-degree: all neurons receive exactly the same number of excitatory and inhibitory connections, $C_E$ and $C_I$, as in \cite{Amit1997, Brunel2000, Mastrogiuseppe2017}. All excitatory synapses have equal strength $J$ and all inhibitory neurons $-gJ$. Furthermore, we consider the large network limit where the number of synaptic neurons $N$ is large while keeping the excitatory and inhibitory inputs $C_E$ and $C_I$ fixed.


\subsection*{Single unit: timescales of dynamics}

\begin{figure}[!h]
 	\begin{adjustwidth}{-1.25in}{0in}
 		\includegraphics[width=\linewidth]{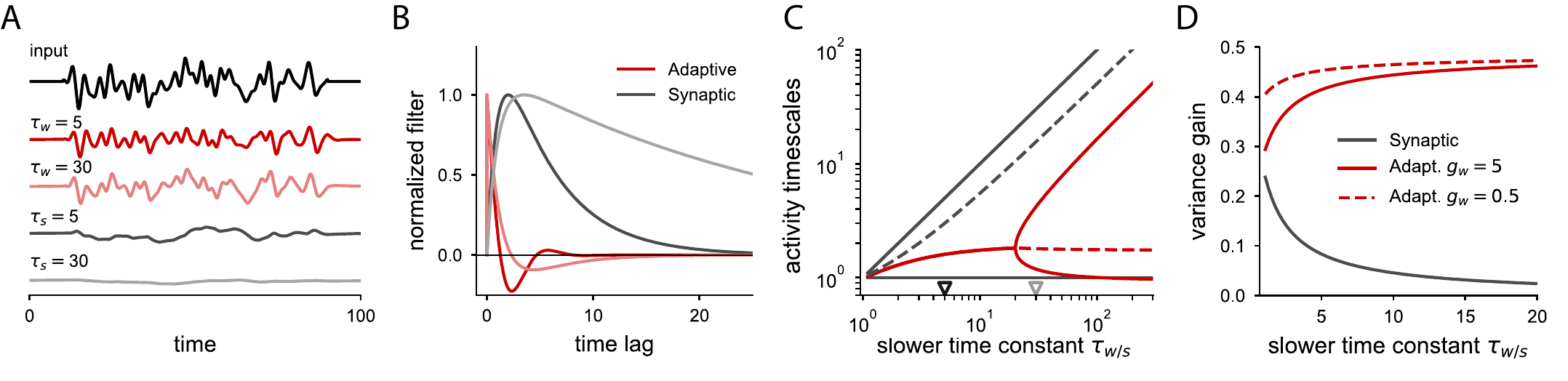}
 		\caption{{\bf Activity of individual neurons with adaptation or synaptic filtering.} A: Firing rate response of two different neurons with adaptation (red curves) and two different neurons with synaptic filtering (grey curves) to the same time-varying input (black curve). B: Normalized linear filters for the neurons shown in A.   C: Timescales of the linear filter for neurons with adaptation (red lines) and for neurons  with synaptic filtering (grey lines) as a function of the timescale $\tau_w$ or $\tau_s$, respectively. The dashed lines indicate the effective timescale of the evoked activity obtained by weighing each individual timescale with its amplitude in the linear filter. The effective timescale for neurons with adaptation saturates for large adaptation time constants, while it grows proportionally to the synaptic time constant for neurons with synaptic filtering. Note that for the adaptive neuron, if the two eigenvalues are complex conjugate, there is only one decay timescale. The triangles on the temporal axis indicated the time constants used in A and B. Adaptation coupling $g_w = 5$. D: Variance of the input current as a function of the slow time constant when the adaptive and synaptic neurons are stimulated with Gaussian white noise of unit variance. In the case of neurons with adaptation, two different values of the adaptation coupling $g_w$ are shown.}
 		\label{fig1}
 	\end{adjustwidth}
 \end{figure}

 In the models studied here the input current of individual neurons is described by a linear system. Thus, their activity is fully characterized by the response $h\left(t\right)$ to a brief impulse signal, i.e. the linear filter. When such neurons are stimulated with a time-varying input $I\left(t\right)$, the response  is the convolution of the filter with the input, $x\left(t\right) = \left(h\ast I \right)\left(t\right)$. These filters can be determined analytically for both neurons with adaptation or synaptic filtering and directly depend on the parameters of these processes. Analyzing the differences that these two slow processes produce in the linear filters is useful for studying the differences in the response of adaptive and synaptic filtering neurons to temporal stimuli (Fig~\ref{fig1} A), and will serve as a reference for comparison to the effects that emerge at the network level.

 In particular, the filter of a neuron with synaptic filtering, $h_s\left(t\right)$, is the sum of two exponentially decaying filters of opposite signs and equal amplitude, with time constants $\tau_s$ and $\tau_m$: 

\begin{equation}
h_s\left(t\right) = \frac{1}{\tau_s-\tau_m} \left(e^{-\frac{t}{\tau_s}}-e^{-\frac{t}{\tau_m}}\right) \Theta\left(t\right),
\end{equation}

\noindent where $\Theta\left(t\right)$ is the Heaviside function (see \textit{Methods}). Thus, the current response of a neuron to an input pulse received from an excitatory presynaptic neuron is positive and determined by two different timescales. The response first grows with timescale $\tau_m$, so that the neuron cannot respond to any abrupt changes in the synaptic input faster than this timescale, and then decreases back to zero with timescale $\tau_s$ (grey curves, Fig~\ref{fig1} B).

The adaptation filter is given as well by the linear combination of two exponential functions. In contrast to the synaptic filter, since the input in the adaptive neuron model affects directly the current variable $x_i\left(t\right)$, there is an instantaneous change in the firing rate to an input delta-function (red curves, Fig~\ref{fig1} B). The timescales of the two exponentials can be calculated as

\begin{equation}
\tau^\pm = \frac{2\tau_m \tau_w}{\tau_w+\tau_m } \left(1\pm \sqrt{1-\frac{4\tau_m \tau_w\left(1+g_w\right)}{\left(\tau_m+\tau_w\right)^2}}\right)^{-1}. \label{tscale_single_syn}
\end{equation}

\noindent When the argument of the square root in Eq~\eqref{tscale_single_syn} is negative, i.e. for an adaptation time constant $\tau_w$ smaller than $\tau_m \left(4\left(1+g_w\right)\right)^{-1}$, the two timescales correspond to a pair of complex conjugate numbers, so that the filter is an oscillatory function whose amplitude decreases monotonically to zero at a single timescale. If the argument of the square root is positive, for slow enough adaptation, the two timescales are real numbers and correspond to exponential functions of opposing signs of decaying amplitude. However, the amplitudes of these two exponentials are different (see \textit{Methods}). To illustrate this, we focus on the limit of large adaptation time constants with respect to the membrane time constant, where the two exponential functions evolve with timescales that decouple the contribution of the membrane time constant and the adaptation current. In that limit,  the adaptive filter reads

\begin{equation}
h_w\left(t\right)  =  \left(-\frac{g_w }{ \tau_w}e^{-\left(1+g_w\right)\frac{t}{\tau_w}}+ \frac{1}{\tau_m}e^{-\frac{t}{\tau_m}}\right) \Theta\left(t\right).\label{adap_filt}
\end{equation}

\noindent The amplitude of the slow exponential is inversely related to its timescale so that the integral of this mode is fixed, and independent of the adaptation time constant. This implies that a severalfold increase of the adaptation time constant does not lead to strong changes in the single neuron activity for time-varying signals (Fig~\ref{fig1}A).

Furthermore, we can characterize the timescale of the single neuron response as the sum of the exponential decay timescales weighed by their relative amplitude, and study how this characteristic timescale evolves as a function of the time constants of either the synaptic or the adaptive current (Fig~\ref{fig1}C). For adaptive neurons, the activity timescale is bounded as a consequence of the decreasing amplitude of the slow mode, i.e. increasing the adaptation time constant beyond a certain value will not lead to a slower response. In contrast, the activity of an individual neuron with synaptic filtering scales proportionally to the synaptic filter time, since the relative amplitudes of the two decaying exponentials are independent of the time constants. 

When any of the two neuron types are stimulated with white Gaussian noise, the variance in the response is always smaller than the input variance, due to the low pass filtering properties of the neurons. However, this gain in the variance of the input currents is modulated by the different neuron parameters (Fig~\ref{fig1}D). For a neuron with synaptic filtering, the gain is inversely proportional to the time constant $\tau_s$. In contrast, for a neuron with adaptation, increasing the adaptation time constant has the opposite effect of increasing the variance of the current response. This is because when the adaptation time constant increases, the amplitude of the slow exponential decreases accordingly, and the low-pass filtering produced by this slow component is weaker. Following the same reasoning, increasing the adaptation coupling corresponds to strengthening the low-pass filtering performed by adaptation, so that the variance decreases (Fig~\ref{fig1}D, dashed vs full red curves). 


\subsection*{Population-averaged dynamics}

\begin{figure}[!h]
		\begin{adjustwidth}{-1.25in}{0in}
		\includegraphics[width=\linewidth]{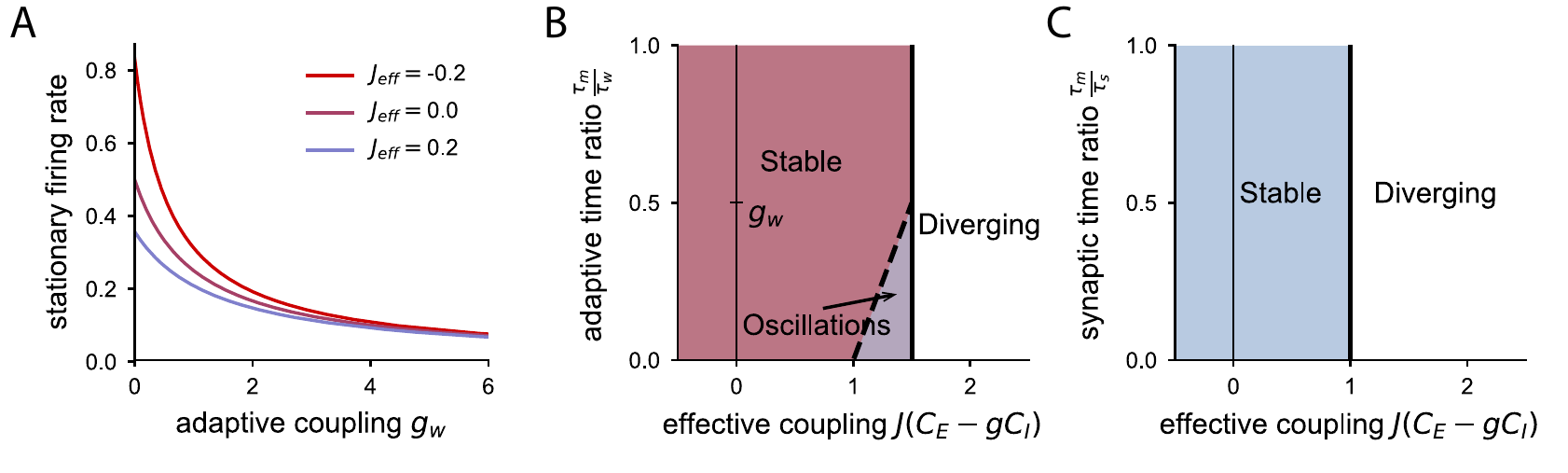}
		\caption{{\bf Equilibrium firing rate and phase diagrams of the population-averaged dynamics.} A: Firing rate of the network with adaptation at the equilibrium $\phi\left(x_0\right)$ for increasing adaptive couplings and three different values of the effective recurrent coupling $J_{\text{eff}} = J\left(C_E-gC_I\right)$. Stronger adaptation leads to lower firing rates at equilibrium. B: Phase diagram of the population-averaged activity for the network with adaptation. C:  Phase diagram for the network with synaptic filtering. }
		\label{fig2}
		\end{adjustwidth}
	\end{figure}

 In the absence of any external input, a non-trivial equilibrium for the population averaged activity emerges due to the recurrent connectivity of the network. The equilibrium firing rate is identical across network units, since all units  are statistically equivalent. We can write the input current $x_0$ at the fixed point as the solution to the transcendental equation
	\begin{equation}
	\left(1+g_w\right)x_0 = J\left(C_E-g C_I\right) \phi \left(x_0\right)+g_w \gamma, \label{trans1}
	\end{equation}
	for the network with adaptation,  and to
	\begin{equation}
	x_0 = J\left(C_E-g C_I\right)  \phi \left(x_0\right), \label{trans2}
	\end{equation}
	 for  synaptic filtering (see \textit{Methods}). Based on Eq~\eqref{trans1}, we find that the adaptation coupling $g_w$ reduces the mean firing rate of the network, independently of whether the network is dominated by inhibition or excitation (Fig~\ref{fig2}A). Synaptic filtering instead does not play any role in determining the equilibrium activity of the neurons, since Eq~\eqref{trans2} is independent of the synaptic filtering parameter $\tau_s$. 
	
	We next study the stability and dynamics of the equilibrium firing rate in response to a small perturbation uniform across the network, $x_i\left(t\right) = x_0 + \delta x\left(t\right)$. Because of the fixed in-degree of the connectivity matrix, the linearized dynamics of each neuron are identical, so that the analysis of the homogeneous perturbation on the network reduces to the study of a two-dimensional deterministic system of differential equations which corresponds to the dynamics of the population-averaged response (see \textit{Methods}). The stability and timescales around equilibrium depend on the two eigenvalues of this linear 2D-system. More specifically, the fixed point is stable to a homogeneous perturbation if the two eigenvalues of the dynamic system have negative real part, in which case the inverse of the unsigned real part of the eigenvalues determines the timescales of the response. For both the network with synaptic filtering and the network with adaptive neurons, the order parameter of the connectivity that determines the stability of the fixed point is the effective recurrent coupling $J\left(C_E-gC_I\right)$ each neuron receives, resulting from the sum of all input synaptic connections. A positive (negative) effective coupling corresponds to a network where recurrent excitation (inhibition) dominates and the recurrent input provides positive (negative) feedback \cite{Brunel2000, Mastrogiuseppe2017}. 
	
	For networks with synaptic filtering, we find that the synaptic time constant does not alter the stability of the equilibrium state, so that the effective coupling alone determines the stability of the population-averaged activity. As the effective input coupling strength is increased, the system undergoes a saddle-node bifurcation when the effective input is $J\left(C_E-gC_I\right) = 1$ (Fig~\ref{fig2}C). In other words, the strong positive feedback loop generated by the excitatory recurrent connections destabilizes the system. 
	
	 
%
 To analyze the timescales elicited by homogeneous perturbations, we calculate the eigenvalues and eigenvectors of the linearized dynamic system (see \textit{Methods}). We find that for inhibition-dominated networks ($J\left(C_E-gC_I\right)<0$), the network shows population-averaged activity at timescales that interpolate between the membrane time constant and the synaptic time constant. As the effective coupling is increased, the slow timescale at the network level can be made arbitrarily slow by tuning the effective synaptic coupling close to the  bifurcation value, a well-known network mechanism to achieve slow neural activity \cite{Huang2017}.
 
 In the limit of very slow synaptic timescale, the two  timescales of the population-averaged activity are
 
 \begin{eqnarray}
 \tau^+ &=& \frac{\tau_s}{1-J\left(C_E-gC_I\right)},\\
 \tau^- &=& \tau_m\left(1-J\left(C_E-gC_I\right)\frac{\tau_s}{\tau_m}\right),  \end{eqnarray} 
	\noindent so that the timescale $\tau^-$ is proportional to the membrane time constant and $\tau^+$ is proportional to the slow synaptic time constant, effectively decoupling the two timescales.  The relative contribution of these two timescales is the same, independently of the time constant $\tau_s$, as we found in the single neuron analysis.

The network with adaptation shows different effects on the population-averaged activity. First, the presence of adaptation modifies the region of stability: the system is stable when the effective recurrent input $J\left(C_E-gC_I\right)$ is less than the minimum of $1+g_w$ and $1+\frac{\tau_m}{\tau_w}$ (see \textit{Methods}). Therefore, the stability region is larger than for the network with synaptic filtering (Fig~\ref{fig2}B vs Fig~\ref{fig2}C). In other words, the effective excitatory feedback required to destabilize the network is larger due to the counterbalance provided by adaptation. Moreover, adaptation allows the network to undergo two different types of bifurcations as the effective input strength increases, depending on the adaptation parameters. One possibility is a saddle-node bifurcation, as in the synaptic case, which takes place when $J\left(C_E-gC_I\right) = 1 + g_w$. Beyond the instability all neurons in the network saturate. The other possible bifurcation, which happens if $\frac{\tau_m}{\tau_w}<g_w$, at an effective coupling strength $J\left(C_E-gC_I\right) = 1+\frac{\tau_m}{\tau_w}$, is a Hopf bifurcation: the fixed point of network becomes unstable, leading in general to oscillating dynamics of the population-averaged response. Note that in the limit of very slow adaptation, the system can only undergo a Hopf bifurcation (Fig~\ref{fig2}B).
	
The two timescales of the population-averaged activity in the stable regime for the adaptive network decouple the two single neuron time constants when adaptation is much slower than the membrane time constant. In this limit, up to first order of the adaptive time ratio $\frac{\tau_m}{\tau_w}$, the two activity timescales are

\begin{align}
\tau^+ &= \frac{\tau_m}{1-J\left(C_E-gC_I\right)},\\
\tau^- &= \frac{\tau_w \left(1-J\left(C_E+gC_I\right)\right)}{1+g_w-J\left(C_E-gC_I\right)}.\label{approxTadap}
\end{align}

\noindent Similar to the single neuron dynamics, the amplitude of the slow mode, corresponding to $\tau^-$, decreases as $\tau_w$ is increased, so that the contribution of the slow timescale is effectively reduced when $\tau_w$ is very large. On the contrary, the mode corresponding to $\tau^+$, proportional to the membrane time constant can be tuned to reach arbitrarily large values. This network mechanism to obtain slow dynamics does not depend on the adaptation properties.

\subsection*{Heterogeneous activity}

	\subsubsection*{Linear stability analysis}
	
\begin{figure}[!h]
		\begin{adjustwidth}{-1.5in}{0in}
		\includegraphics[width=\linewidth]{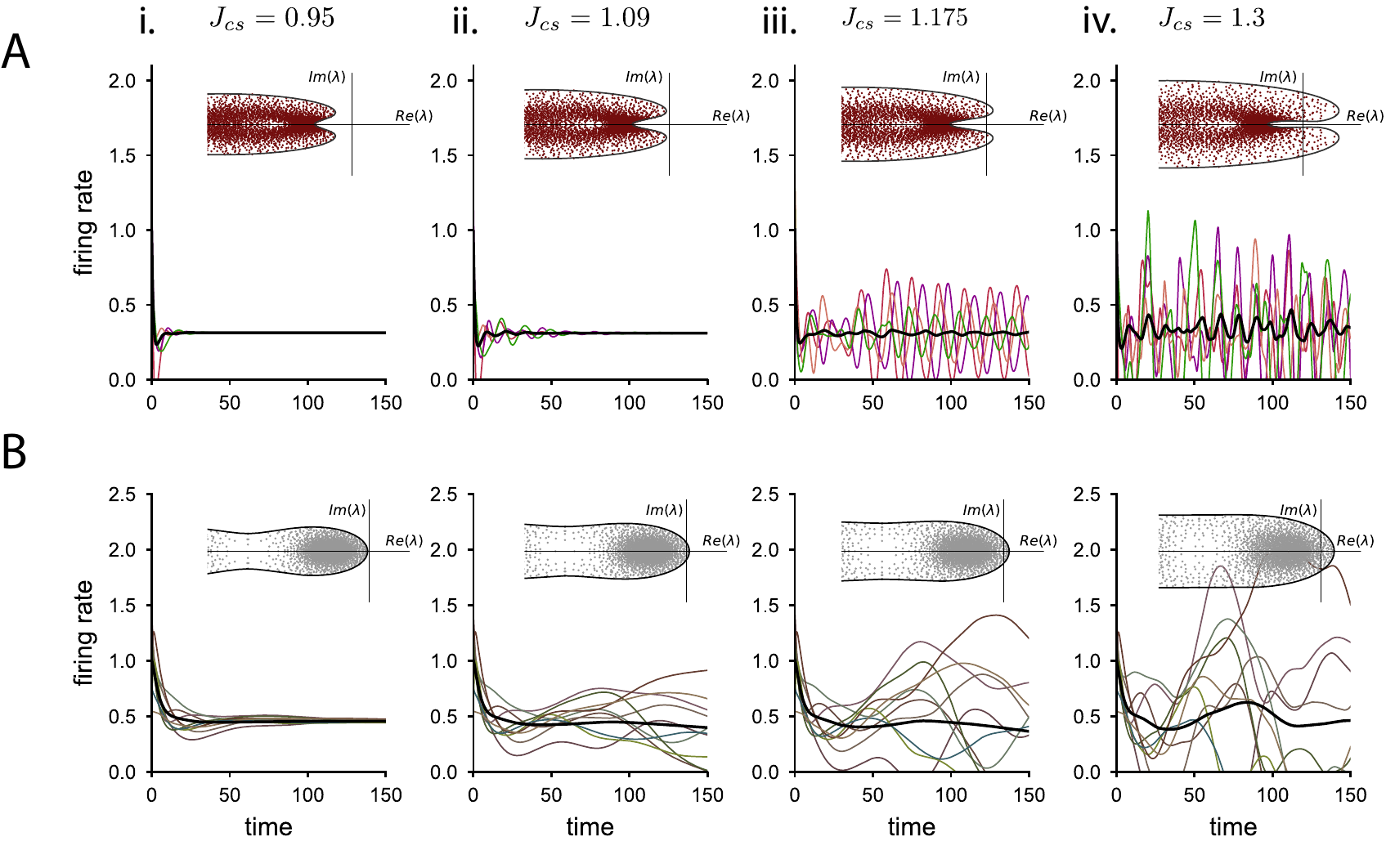}
		\caption{{\bf Dynamical regimes as the coupling strength is increased.} Numerical integration of the dynamics for the network with adaptive neurons (row A) and the network with synaptic filtering (row B), as the coupling variance $J_{cs} = J\sqrt{C_E+g^2C_I}$ is increased. Colored lines correspond to the firing rates of individual neurons, the black line indicates the population average activity. Insets: complex eigenspectrum $\lambda_{w/s}$ of the linearized dynamical matrix around the fixed point. Dots: eigenvalues of the connectivity matrix used in the network simulation. Solid line: theoretical prediction for the envelope of the eigenspectrum.  The imaginary axis, $\text{Re}\left(\lambda\right) =0$, is the stability boundary. i. Both the network with adaptation and synaptic transmission are stable. ii. The network with synaptic filtering crosses the stability boundary and shows fluctuations in time and across neurons, while the network with adaptation remains stable. iii. The network with synaptic filtering displays stronger fluctuations. The network with adaptive neuron undergoes a Hopf bifurcation leading to strong oscillations at a single frequency with uncorrelated phases across units. Note in the inset that for this connectivity matrix there is only one pair of complex conjugate unstable eigenvalues in the finite network. iv. The network with synaptic filtering shows strong fluctuations. The network with adaptation displays fluctuating activity with an oscillatory component. Parameters: in A, $g_w=0.5$, and $\tau_w = 5$, in B, $\tau_s=5$.}
		\label{fig3}
		\end{adjustwidth}
	\end{figure}

 Previous studies have shown that random connectivity can lead to heterogeneous dynamics where the activity of each unit fluctuates strongly in time \cite{Sompolinsky1988, Rajan2010, Kadmon2015, Mastrogiuseppe2017}. To assess the effects of additional hidden degrees of freedom on the emergence and timescales of such fluctuating activity, we examine the dynamics when each unit is perturbed independently away from the equilibrium, $x_i\left(t\right) = x_0 + \delta x_i \left(t\right)$. By linearizing the full $2N$-dimensional dynamics around the fixed point, we can study the stability and timescales of the activity characterized by the set of eigenvalues of the linearized system, $\lambda_s$ and $\lambda_w$ for the network with synaptic filtering neurons and adaptation, respectively. These sets of eigenvalues are determined by a direct mapping to the eigenvalues of the connectivity matrix, $\lambda_J$ (see \textit{Methods}). The eigenvalues $\lambda_J$ of the connectivity matrices considered are known in the limit of large networks \cite{Rajan2006, Mastrogiuseppe2017}: they are enclosed in a circle of radius $J\sqrt{C_E+g^2 C_I}$, except for an outlier that corresponds to the population-averaged dynamics, studied in the previous section. Therefore, we can map the circle that encloses  the eigenspectrum $\lambda_J$ into a different shape in the space of eigenvalues $\lambda_{s/w}$ (insets Fig. \ref{fig3}). In order to determine the stability of the response to the perturbation, we assess whether the real part of the eigenspectrum $\lambda_{s/w}$ is negative at all possible points. Furthermore, the type of bifurcation is determined by whether the curve enclosing the eigenvalues $\lambda_{s,w}$ crosses the imaginary axis at zero frequency or at a finite frequency when the synaptic coupling strength is increased, leading respectively to a zero-frequency or to a Hopf bifurcation \cite{Bimbard2016}.
	
	The order parameter of the connectivity that affects the stability and dynamics of the network is now the radius of the circle of eigenvalues $\lambda_J$, i.e. $J\sqrt{\left(C_E + g^2 C_I\right)}$. This parameter is the standard deviation of the synaptic input weights of a neuron (see \textit{Methods}), which contrasts with the order parameter of the population-averaged response, that depends on the mean of the synaptic input weights. The mean and standard deviation of the synaptic connectivity can be chosen independently, so that while the population-averaged activity remains stable, the individual neurons might not display stable dynamics. To analyze solely the heterogeneous response of the network to the perturbation, we focus in the following on network connectivities whose population-averaged activity is stable, i.e. the effective synaptic coupling is inhibitory or weakly excitatory.
	
\begin{figure}[!h]
		\begin{adjustwidth}{-2.25in}{0in}
			\includegraphics[width=\linewidth]{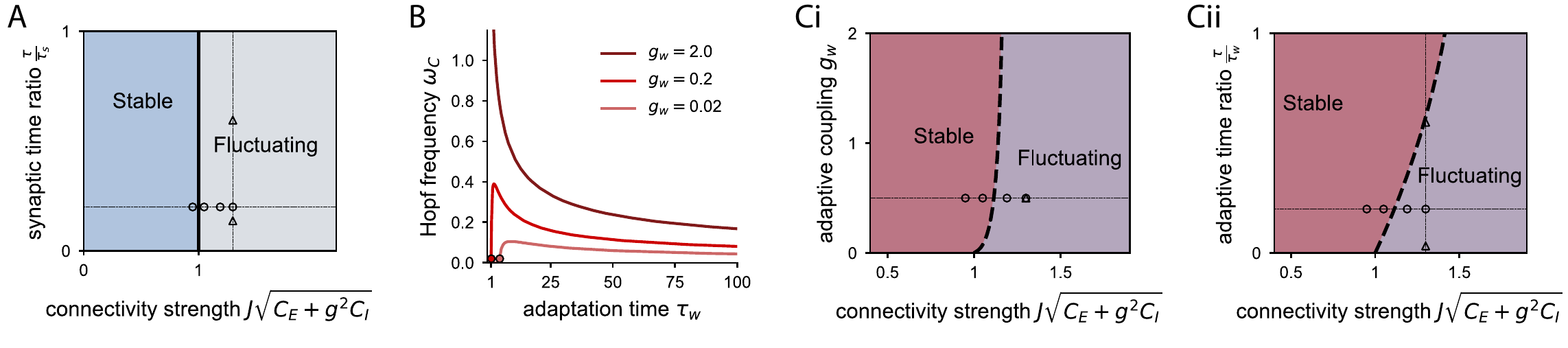}
			\caption{\textbf{Phase diagram and frequency of the bifurcation for the heterogeneous activity.}
				A: Phase diagram for the network with synaptic transmission. The only relevant parameter to assess the dynamical regime is the connectivity strength. The circles indicate the parameters used in Figs~\ref{fig3} and \ref{fig6}. Triangles correspond to the parameter combinations used in Fig~\ref{fig5}. B: Frequency at which the eigenspectrum loses stability for the network with adaptive neurons as a function of the ratio between membrane and adaptation time constant, $\tau_m/\tau_w$, for three different adaptive couplings. The dots indicate the fastest adaptive time constant for which the system undergoes a Hopf bifurcation (Eq.~\ref{sadHopf}). C: Phase diagrams for the two adaptation parameters, (i) the coupling $g_w$ and (ii) the adaptive time constant $\tau_w$ vs the coupling strength. }
			\label{fig4}
		\end{adjustwidth}
	\end{figure}

 We find that in the network with synaptic filtering, the eigenspectrum $\lambda_s$ always crosses the stability bound through the real axis, which takes place when the spectral radius of the connectivity is one, $J\sqrt{C_E+g^2 C_I} = 1$. Thus the system undergoes a zero-frequency bifurcation similar to randomly connected networks without hidden variables \cite{Sompolinsky1988, Kadmon2015, Schuecker2017, Mastrogiuseppe2017}, leading to strong fluctuations at the single neuron level that are self-sustained by the network connectivity (Fig~\ref{fig3} Bii-Biv). The critical coupling at which the equilibrium firing rate loses stability is independent of the synaptic time constant, i.e. synaptic filtering does not affect the stability of heterogeneous responses (Fig~\ref{fig4} A). However, the synaptic time constant $\tau_s$ affects the timescales at which the system returns to equilibrium after a perturbation, because the eigenvalues $\lambda_s$ (see Eq~\eqref{map_syn} in \textit{Methods}) depend explicitly on $\tau_s$. 
	 
For a network with adaptive neurons, we calculate the eigenspectrum $\lambda_w$ and find that the transition to instability $\text{Re} \left(\lambda_w\right) = 0$ can happen either at zero frequency
	or at a finite frequency (see \textit{Methods}), leading to a Hopf bifurcation (as in inset Fig~\ref{fig3} Aiii). In particular, the network dynamics undergo a Hopf bifurcation when
	\begin{equation}
 \tau_w>  \frac{ \tau_m}{{g_w+\sqrt{2g_w\left(g_w+1\right)}}},
	\end{equation}
	
	\noindent so that strong adaptation coupling and slow adaptation time constants lead to a finite frequency bifurcation. In particular, if the coupling $g_w$ is larger than $\sqrt{5}-2\approx 0.236$, only the Hopf bifurcation is possible, since by construction $\frac{\tau_m}{\tau_w}<1$. We can also calculate the frequency of oscillations at the Hopf bifurcation. We find that, for slow adaptive currents, the Hopf frequency is inversely related to the adaptation time constant (Fig~\ref{fig4}B), so that slower adaptation currents produce slower oscillations at the bifurcation. 
	
	Adaptation also increases the stability of the equilibrium firing rate to a heterogeneous perturbation, in comparison to a network with synaptic filtering (Fig~\ref{fig4} C). This can be intuitively explained in geometrical terms by analyzing how adaptation modifies the shape of the eigenspectrum $\lambda_w$ with respect to the circular eigenspectrum of the connectivity matrix $\lambda_J$.

	The Hopf bifurcation leads to the emergence of a new dynamical regime in the network (Fig~\ref{fig3} Aiv), which is studied in the following section. Right at the Hopf bifurcation, the system shows marginal oscillations at a single frequency that can be reproduced in finite-size simulations whenever only one pair of complex conjugate eigenvalues is unstable (Fig~\ref{fig3} Aiii). 
	
	\subsubsection*{Fluctuating activity: dynamical mean field theory}
	
	The classical tools of linear stability theory applied so far can only describe the dynamics of the system up to the bifurcation. To study the fluctuating regime, we take a different approach and focus on the temporal statistics of the activity, averaged over different connectivity matrices: we determine the mean and autocorrelation function of the single neuron firing rate, and characterize the timescale of the fluctuating dynamics \cite{Sompolinsky1988, Rajan2010, Aljadeff2015, Harish2015, Kadmon2015, Schuecker2017, Mastrogiuseppe2017}. For large networks, the dynamics can be statistically described by applying dynamical mean field theory (DMFT), which approximates the deterministic input to each unit by an independent Gaussian noise process. The full network is then reduced to a two-dimensional stochastic differential equation, where the first and second moments of the noise must be calculated self-consistently. We solve the self-consistent equations using a numerical iterative procedure, similar to the schemes followed in \cite{Stern2014, Lerchner2006a, Dummer2014, WieBer15, Rajan2010} (see \textit{Methods} for an explanation of the iterative algorithm and its practical limitations).

\begin{figure}[!h]
				\begin{adjustwidth}{-1.75in}{0in}
				\includegraphics[width=\linewidth]{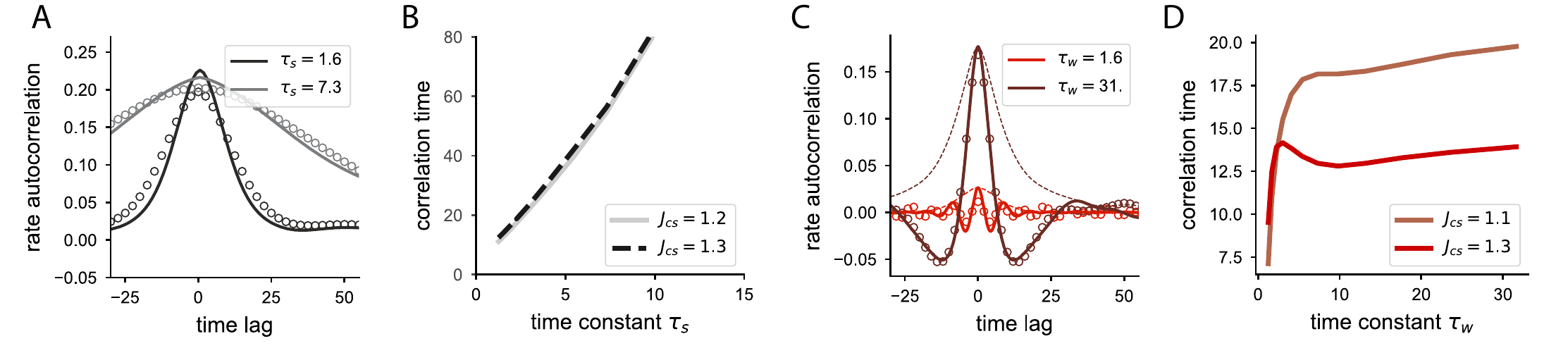}
				\caption{{\bf Autocorrelation function and timescale of the network activity in the fluctuating regime.} A: Autocorrelation function of the firing rates in the network with synaptic filtering; dynamical mean field results (solid lines) with their corresponding envelopes (dashed lines), and results from simulations (empty dots). Connectivity strength $J_{cs} = J\sqrt{C_E+g^2 C_I} = 1.2$. B: Effective timescale of the network activity as a function of the synaptic time constant for the network with synaptic filtering. The network coupling does not have a strong effect on the effective timescale. C: Autocorrelation function of the firing rates, as in A, for the system with adaptive neurons. $J_{cs} = 1.3$. D: Effective timescale of the firing rates, as in B, for the system with adaptive currents. }
				\label{fig5}
			\end{adjustwidth}
			\end{figure}

	For the network with synaptic filtering, we find that the autocorrelation function of the firing rates in the fluctuating regime corresponds to a monotonically decreasing function (Fig~\ref{fig5} A), qualitatively similar to the correlation obtained in absence of synaptic filtering \cite{Mastrogiuseppe2017}. This fluctuating state has often been referred to as rate chaos and shows non-periodical heterogeneous activity which is intrinsically generated by the network connectivity. The main effect of synaptic filtering is on the timescale of these fluctuations. When the synaptic time constant is much larger than the membrane time constant, the timescale of the network activity is proportional to the synaptic time constant $\tau_s$, as indicated by the linear dependence between the half-width of the autocorrelation function and the synaptic timescale $\tau_s$, when all other network parameters are fixed (Fig~\ref{fig5} B). 
		
		 For the network with adaptation, we focus on large adaptation time constant $\tau_w$, where the network dynamics always undergo a Hopf bifurcation. The autocorrelation function in such a case displays damped oscillations (Fig~\ref{fig5} C). The decay in the envelope of the autocorrelation function is due to the chaotic-like fluctuations of the firing rate activity. 
		 
		 We define the time lag at which the envelope of the autocorrelation function decreases as the timescale of the network dynamics (see \textit{Methods}). The timescale of the activity increases as the adaptation timescale is increased, when all the other parameters are fixed (Fig~\ref{fig5} D). However, this activity timescale saturates for large values of the adaptation timescale: the presence of very slow adaptive currents, beyond a certain value, will not slow down strongly the network activity. This saturation value depends on the connectivity strength.
		  
	\paragraph{Effects of noise}
	
	 The networks studied so far, for a fixed connectivity matrix, are completely deterministic. We next study the effects of additional white noise inputs to each neuron, as a proxy towards understanding recurrent networks of spiking neurons with adaptation and synaptic filtering. On the mean-field level, such noise is equivalent to studying a recurrent network whose neurons fire action potentials as a Poisson process with instantaneous firing rate $\phi\left(x_i\left(t\right)\right)$ \cite{Ostojic2011, Kadmon2015}. 
	 
\begin{figure}[!h]
	 	\begin{adjustwidth}{-1.25in}{0in}
	 		\includegraphics[width=\linewidth]{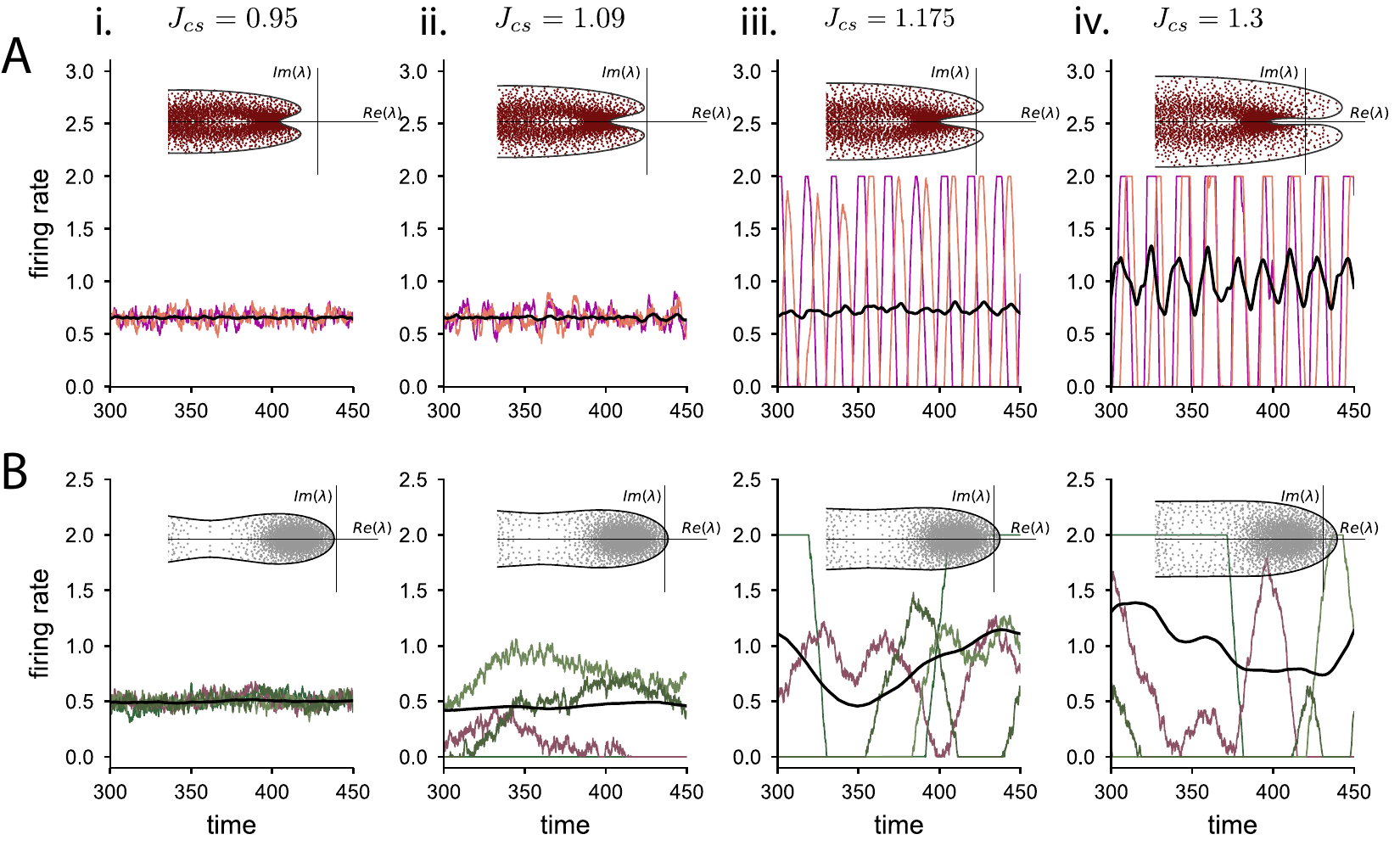}
	 		\caption{{\bf Dynamical regimes for the network with adaptation or synaptic filtering with additive external noise.} Numerical integration of the dynamics for the network with adaptive neurons (row A) and the network with synaptic filtering (row B) with units receiving additive external white noise, as a proxy for spiking noise. Colored lines correspond to the firing rate of individual neurons, the black line indicates the population average activity. Insets: complex eigenspectrum $\lambda_{w/s}$ of the dynamic matrix at the fixed point. Dots: eigenvalues of the connectivity matrix used in the network simulation. Solid line: theoretical prediction for the envelope of the eigenspectrum.  i. Both the network with adaptation and synaptic transmission are stable, the external noise generates stationary fluctuations around the fixed point. ii. The network with synaptic filtering undergoes a zero-frequency bifurcation. Noise adds fast temporal variability in the firing rates. The network with adaptation remains stable, and the fluctuations are larger in amplitude. iii. The network with adaptation undergoes a Hopf bifurcation. The firing rate activity combines both the fast fluctuations produced by white noise and the chaotic activity with an oscillatory component. iv. The network with adaptation shows highly irregular activity, and strong effects due to the activation and saturation bounds of the transfer function. Parameters as in Fig~\ref{fig4}, external noise $\sigma_{\eta}= 0.06$.}
	 		\label{fig6}
	 	\end{adjustwidth}
	 \end{figure}

	 Numerical simulations show that in the stable regime the additive external noise generates weak, fast stationary dynamics around the fixed point (Fig~\ref{fig6} Ai, Bi). The timescale of these fluctuations and their amplitude depend on the distance of the eigenspectrum to the stability line, so that the stable fluctuations for weak synaptic coupling strength (Fig~\ref{fig6}Ai) are smaller in amplitude than those for larger coupling strength (Fig~\ref{fig6}Aii), whose eigenspectrum is closer to the stability boundary. For adaptation, in the fluctuating regime beyond the Hopf bifurcation, the network activity shows again a combination of fluctuating activity and oscillations. 
	 
\begin{figure}[!h]
	\begin{adjustwidth}{-1.25in}{0in}
		\includegraphics[width=\linewidth]{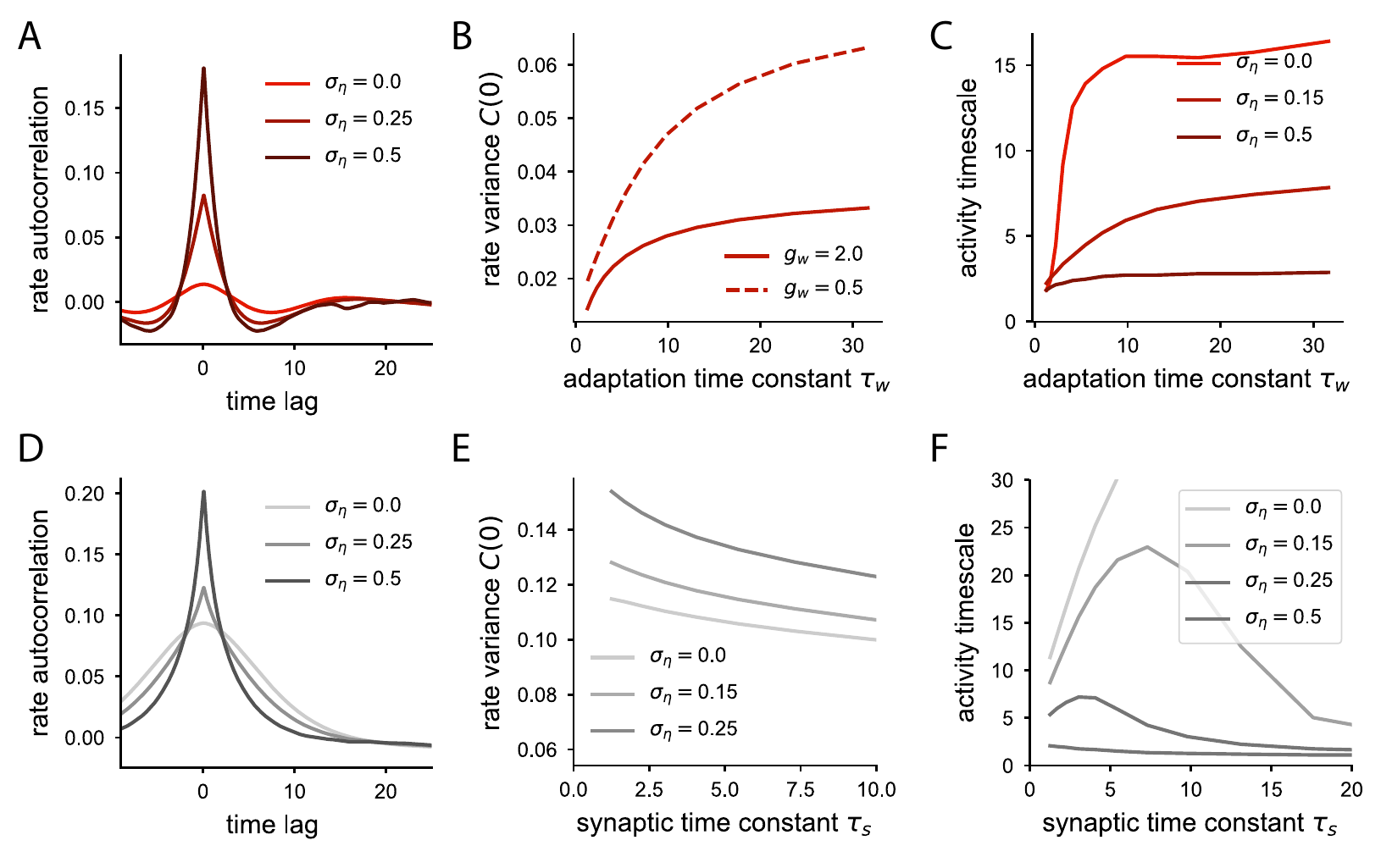}
		\caption{\textbf{Autocorrelation function, variance of the firing rates and timescale of the network activity with external noise predicted by dynamical mean field theory.} A: Autocorrelation function of the firing rates for the network with adaptive neurons for three different noise intensities. Adaptation time constant $\tau_w = 1.25$. B: Variance of the firing rate as a function of the adaptation time constant for two different adaption couplings $g_w$. Increasing the adaptation time constant or decreasing the adaptation coupling increases the variance. $\sigma_\eta = 0.15$. C: Timescale of the firing rate as a function of the adaptation time constant, and three different noise levels. Parameters: $g_w = 0.5$, and $J\sqrt{C_E+g^2 C_I} = 1.2$. D: Autocorrelation function of the firing rate for the network with synaptic transmission for three different noise levels. Synaptic time constant $\tau_s = 1.25$. E: Variance of the firing rate as a function of the synaptic time constant, for three different external noise levels. Synaptic filtering reduces the variance. F: Timescale of the activity for the network with synaptic filtering and external noise. }
		\label{fig7}
	\end{adjustwidth}
\end{figure}

	We further extend the DMFT analysis to account for the additional variance of the external white noise sources (see \textit{Methods}). The autocorrelation function of the firing rates, as predicted by DMFT, does not vary drastically when weak noise is added to the network, except for very short time lags, at which white noise introduces fast fluctuations (see Fig~\ref{fig7}). For the network with adaptation, the autocorrelation function of the firing rates still shows damped oscillations (Fig~\ref{fig7} A), while for the network with synaptic filtering, similarly, weak noise does not affect much the decay of the autocorrelation function (Fig~\ref{fig7} D). Very strong external noise on the other hand will reduce the effect of the underlying recurrent dynamics of the rate network, since the signal to noise ratio in the synaptic input of all neurons is low.

For a fixed external noise intensity, reducing the adaptation coupling or increasing the adaptation time constant increases the variance of the firing rate (Fig~\ref{fig7}B), which resembles the dependence of the variance gain for individual neurons (Fig~\ref{fig1}D). Conversely, slower synaptic filtering reduces the variance of the neuron's firing rates. This is because in the network with synaptic filtering the noise is also filtered at the synapses --in the limit of very large $\tau_s$, the whole white noise is filtered out-- whereas in the network with adaptation the noise affects directly the input current, without being first processed by the adaptation variable. 

However, the timescale of the activity is nonetheless drastically affected by strong noise. External noise adds fast fluctuations on top of the intrinsically generated dynamics of the heterogeneous network with adaptation or synaptic filtering. If the noise is too strong, the effective timescale of the activity takes into account mostly this fast component. In that limit, the timescale of the activity is almost independent of the synaptic or adaptive time constants (Fig~\ref{fig7} C and F, largest noise intensity).

\section*{Discussion}

We examined dynamics of excitatory-inhibitory networks in which each unit had a hidden degree of freedom that represented either firing-rate adaptation or synaptic filtering. The core difference between adaptation and synaptic filtering was how external inputs reached the single-unit activation variable that represents the membrane potential. In the case of adaptation, the inputs directly entered the activation variable, which was then filtered by the hidden, adaptive variable through a negative feedback loop. In the case of synaptic filtering, the external inputs instead reached first the hidden, synaptic variable and were therefore low-pass filtered before being propagated in a feed-forward fashion to the activation variable. While both mechanisms introduce a second timescale in addition to the membrane time constant, our main finding is that the interplay between those two timescales is very different in the two situations. Surprisingly, in presence of adaptation, the membrane timescale remains the dominant one in the dynamics, while the contribution of the adaptation timescale appears to be weak.
In contrast, in a network with synaptic filtering, the dominant timescale of the dynamics is directly set by the synaptic variable, and the overall dynamics are essentially equivalent to a network in which the membrane time-constant is replaced with the synaptic one.

We used a highly abstracted model, in which each neuron is represented by membrane current that is directly transformed into a firing-rate through a non-linear transfer function. This class of models has been popular for dissecting dynamics in excitatory-inhibitory \cite{Wilson1972, Wilson1973, Troyer1997,  Murphy2009, Ahmadian2013} or randomly-connected networks \cite{Sompolinsky1988, Abbott1994, Mastrogiuseppe2017}, and for implementing computations \cite{Jaeger2001, Sussillo2009}. Effects of adaptation in this framework have to our knowledge not been examined so far. We therefore extended the standard rate networks by introducing adaptation in an equally abstract fashion\cite{Benda2003}, as a hidden variable specified solely by a time constant and a coupling strength. Different values of those parameters can be interpreted as corresponding to different  specific membrane conductances that implement adaptation, e.g. the calcium dependent potassium $I_{ahp}$ current or the slow voltage-dependent potassium current $I_m$, which are known to exhibit timescales over several orders of magnitude \cite{Brown2000, Stanley2011}. To cover the large range of adaptation timescales observed in experiments \cite{LaCamera2006}, it would be straightforward to superpose several hidden variables with different time constants. Our approach could also be easily extended to include simultaneously adaptation and synaptic filtering.

A number of previous works have studied the effects of adaptation within more biologically constrained, integrate-and-fire models. These works have in particular examined the effects of adaptation on the spiking statistics \cite{Naud2008,Schwalger2010, Ladenbauer2014}, firing-rate response \cite{Richardson2003, Brunel2003}, synchronisation \cite{Ermentrout2001, Ladenbauer2012, Augustin2013, Ladenbauer2014, Schwalger2013}, perceptual bistability \cite{Laing2002} or single-neuron coding \cite{Naud2012,Pozzorini2013}. In contrast, we have focused here on the relation between the timescales of adaptation and those of network dynamics. While our results rely on a simplified firing-rate model, we expect that they can be directly related to networks of spiking neurons by exploiting quantitative techniques for mapping adaptive integrate-and-fire models to effective firing rate descriptions\cite{augustin2017}.

A side result of our analysis is the finding that strong coupling in random recurrent networks with adaptation generically leads to a novel dynamical state, in which individual units exhibit a mixture of oscillatory and strong temporal fluctuations. The characteristic signature of this dynamical state is a damped oscillation found in the auto-correlation function of single-unit activity. In contrast, classical randomly connected networks lead to a fluctuating, chaotic state in which the auto-correlation function decays monotonically \cite{Sompolinsky1988, Rajan2010, Kadmon2015, Mastrogiuseppe2017}. Note that the oscillatory activity of different units is totally out of phase, so that no oscillation is seen at the level of population activity. This dynamical phenomenon is analogous to heterogeneous oscillations in anti-symmetrically connected networks with delays \cite{Bimbard2016}. In both cases, the oscillatory dynamics emerge through a bifurcation in which a continuum of eigenvalues crosses the instability line at a finite-frequency. Similar dynamics can be also found in networks in which the connectivity is a superposition of a random and a rank two structured part \cite{Mastrogiuseppe2017}. In that situation, the heterogeneous oscillations however originate from a Hopf bifurcation due to an isolated pair of eigenvalues that correspond to the structured part of the connectivity.

Our main aim here was to determine how hidden variables could induce long timescales in randomly-connected networks. Long timescales could alternatively emerge from non-random connectivity structure. As extensively investigated in earlier works, one general class of mechanism relies on setting the connectivity parameters close to a bifurcation that induces arbitrarily long timescales \cite{Sompolinsky1988, Huang2017}. Another possibility is that non-random features of the connectivity, such as the over-representation of reciprocal connections \cite{Sjostrom2001, Ko2011} slow down the dynamics away from any bifurcation. A recent study \cite{Marti2018} has indeed found such a slowing-down. Weak connectivity structure of low-rank type provides yet another mechanism for the emergence of long timescales. Indeed, rank-two networks can generate slow manifolds corresponding to ring attractors provided a weak amount of symmetry is present \cite{Mastrogiuseppe2018}.

Ultimately, the main reason for looking for long timescales in the dynamics is their potential role in computations performed by recurrent networks \cite{Sussillo2014,Barak2017}. Recent works have proposed that adaptive currents may help implement computations in spiking networks by either introducing slow timescales or reducing the amount of noise due to spiking \cite{Nicola2016,Bellec2018}.  Our results suggest that synaptic filtering is a much more efficient mechanism to this end than adaptation. Identifying a clear computational role for adaptation in recurrent networks therefore remains an open and puzzling question.

\section*{Methods}
\subsection*{Network model}

We compare the dynamics of two different models: a recurrent network with adaptive neurons, and a recurrent network with synaptic filtering.   Each model is defined as a set of $2N$ coupled differential equations. The state of the $i$-th neuron is determined by two different variables, the input current $x_i\left(t\right)$ and the adaptation (synaptic) variable $w_i\left(t\right)$ ($s_i\left(t\right)$). 

\paragraph{Adaptation } The dynamics of the recurrent network with adaptive neurons are given by 
 
	\begin{equation}
	\begin{cases}
	&\tau_m \, \dot{x}_i \left(t\right)=-x_i\left(t\right) -g_w w_i\left(t\right)+I_i\left(t\right)\\
	& \tau_w \dot{w}_i\left(t\right) = -w_i\left(t\right) + \phi\left(x_i\left(t\right)\right),
	\end{cases} \label{Adap_r}
	\end{equation}
	
	\noindent where $\phi\left(x\right)$ is a monotonically increasing non-linear function that transforms the input current into firing rate. In this study, we use a threshold-linear transfer function with saturation:
	
	\begin{equation}
	\phi\left(x\right) = \begin{cases}
	\left[x-\gamma\right]^+ &\text{if } x-\gamma <\phi_{\text{max}} \\
	\phi_{\text{max}}  &\text{otherwise}.
	\end{cases} \label{transfer_met}
	\end{equation}
	
	In Eq~\eqref{Adap_r} adaptation in single neuron rate models is defined as a low-pass filtered version with timescale $\tau_w$ of the neuron's firing rate $\phi\left(x_i\left(t\right)\right)$, and is fed back negatively into the input current, with a strength that we call the adaptation coupling $g_w$. For the sake of mathematical tractability, we linearize the dynamics of the adaptation variable by linearizing the transfer function (Eq~\ref{transfer_met}),  $\phi\left(x_i\left(t\right)\right) \approx x_i\left(t\right) - \gamma$.  Therefore, the dynamics of the network model with adaptation studied here read
	
	\begin{equation}
	\begin{cases}
	&\tau_m \, \dot{x}_i \left(t\right)=-x_i\left(t\right) -g_w w_i\left(t\right)+I_i\left(t\right)\\
	& \tau_w \dot{w}_i\left(t\right) = -w_i\left(t\right) + x_i\left(t\right) - \gamma,
	\end{cases} \label{Adap}
	\end{equation}
	
	Note that this approximation allows for adaptation to increase the input current of a neuron, when the neuron's current is below the activation threshold $\gamma$.

	\paragraph{Synaptic filtering} For the recurrent network with synaptic filtering, the dynamics are
	\begin{equation}
	\begin{cases}
	&\tau_m \, \dot{x}_i \left(t\right)=-x_i\left(t\right) +s_i\left(t\right)+I_i\left(t\right)\\
	& \tau_s \dot{s}_i\left(t\right) = -s_i\left(t\right) + I_i\left(t\right).
	\end{cases} \label{Synap}
	\end{equation}

  \noindent In Eqs~\eqref{Adap_r}, \eqref{Adap}, and~\eqref{Synap},  $I\left(t\right)$ represents the total external input received by the neuron. In general, we are interested in the internally generated dynamical regimes of the network, so that the input is given by the synaptic inputs 
  
  \begin{equation}
  I_i\left(t\right) = I_{\text{syn},i} =  \sum_j J_{ij} \phi\left(x_j\left(t\right)\right).
  \end{equation}

  \noindent The matrix element $J_{ij}$ indicates the coupling strength of the $j$-th neuron onto the $i$-th neuron. The connectivity matrix is sparse and random, with constant in-degree \cite{Brunel2000, Ostojic2014, Mastrogiuseppe2017}: all neurons receive the same number of input connections $C$, from which $C_E$ are excitatory and $C_I$ inhibitory. All excitatory synapses have coupling strength $J$ while the strength of all inhibitory synapses is $-gJ$. Moreover, each neuron can only either excite or inhibit the rest of the units in the network, following Dale's principle. Therefore, the total effective input coupling strength, which is the same for all neurons, is
  \begin{equation}
  J_{\text{eff}}:= \sum_j J_{ij} = J\left(C_E-gC_I\right).\label{eff_input}
  \end{equation}
  
  \begin{table}[!ht]
  	\begin{adjustwidth}{0in}{0in} 
  		\centering
  		\caption{
  			{\bf Parameter values.} Parameter values used in the simulations. }
  		\begin{tabular}{|l|l|}
  			\hline
			{\bf Parameter} & {\bf Value} \\ \thickhline
			Number of units $N$ & 3000 \\ \hline
  			In-degree $C$ & 100 \\ \hline
  			Excitatory inputs $C_E$ & 80 \\ \hline
  			Inhibitory inputs $C_I$ & 20 \\ \hline
  			Ratio I-E coupling strength $g$ & 4.1 \\ \hline
  			Threshold $\gamma$ & -0.5 \\ \hline
  			Maximum firing rate $\phi_\text{max}$ & 2\\ \hline
  		\end{tabular}
  		\label{table1}
  	\end{adjustwidth}
  \end{table}

\subsection*{Single neuron dynamics}

 The  dynamics of each individual neuron are described by a two-dimensional linear system, which implies that the input current response $x\left(t\right)$ to a time-dependent input $I\left(t\right)$ is the convolution of the input with a linear filter $h\left(\tau\right)$ that depends on the parameters of the linear system:
 
 \begin{equation}
 x\left(t\right) = \left(h\ast I \right)\left(t\right) = \int_{-\infty}^{+\infty} dt^\prime h\left(t^\prime\right) I\left(t-t^\prime\right)\label{convol}.
 \end{equation}
 
  In general, for any linear dynamic system $\dot{z}\left(t\right) = Az + b\left(t\right)$, where $A$ is a square matrix in $\mathbb{R}^{N\times N}$ and $b\left(t\right)$ is a $N$-dimensional vector, the dynamics are given by
 	
 	\begin{equation}
 	z\left(t\right) = \int_{-\infty}^{\infty} dt^\prime e^{At^\prime} \Theta\left(t^\prime\right)b\left(t-t^\prime\right), \label{def_lsys}
 	\end{equation}
 	
 	\noindent where $\Theta\left(t\right)$ is the Heaviside function. Thus, comparing Eqs \eqref{def_lsys} and \eqref{convol}, the linear filter is determined by the elements of the so-called propagator matrix $P\left(t\right) = e^{At} \Theta\left(t\right)$.
 	
	\paragraph{Synaptic filtering} For a single neuron wit synaptic filtering, the dynamics are given by Equation~\eqref{Synap}, where the input $I_i\left(t\right)$ represents the external current. We write the response in its vector form $ \left(x\left(t\right), s\left(t\right)\right)^T$ and the input as $\left(0, I\left(t\right) \right)^T$. The dynamic matrix is
	\begin{equation}
	A_s = \left(\begin{array}{c c}
	-\tau_m^{-1}  & \tau_m^{-1} \\
	0 & -\tau_s^{-1}
	\end{array}\right).\label{dyn_syn}
	\end{equation}
	
	The linear filter, $h_s\left(t^\prime\right)$, is given by the entries of the propagator matrix that links the input $I\left(t\right)$ to the output element $x\left(t\right)$, which are in this case only the entry in row one and column two: $h_s\left(t^\prime\right) = \left[P\left(t^\prime\right)\right]_{12}$. To compute the required entry of the propagator, we diagonalize the dynamic matrix $A = VDV^{-1} $. The matrix $D$ is a diagonal matrix with the eigenvalues of matrix $A$ in the diagonal entries, and $V$ is a matrix whose columns are the corresponding eigenvectors. Applying the identity $e^{tVDV^{-1}} = Ve^{tD} V^{-1}$ and the definition of propagator we obtain that
	
	\begin{equation}
		h_s\left(t\right) =  \Theta\left(t\right) \frac{1}{\tau_m-\tau_s} \left(e^{-\frac{t}{\tau_m}}-e^{-\frac{t}{\tau_s}}\right).\label{synfilt}
	\end{equation}
	The two timescales of the activity are defined by the inverse of the eigenvalues of the system, which coincide with $\tau_m$ and $\tau_s$. Every time a pulse is given to the neuron, both modes get activated with equal amplitude and opposing signs, as indicated by Eq~\ref{synfilt}. This means that there is a fast ascending phase after a pulse, at a temporal scale  $\tau_m$, and a decay towards zero with timescale $\tau_s$.

	 \paragraph{Adaptation}	The dynamics of a single adaptive neuron are determined by Equation~\eqref{Adap}, where $I_i\left(t\right)$ is the external input to the neuron. We apply the same procedure to determine the timescales of the response of an adaptive neuron to time-dependent perturbations. The dynamic matrix for an adaptive neuron reads
    \begin{eqnarray}
    A_w = \left(\begin{array}{c c}
    -\tau_m^{-1}  & -g_w \tau_m^{-1} \\
    \tau_w^{-1} & -\tau_w^{-1}
    \end{array}\right).\label{dyn_adap}
    \end{eqnarray}
    Its eigenvalues are 	
    \begin{equation}
    \lambda_w^\pm = \frac{1}{2}\left(-\tau_m^{-1}-\tau_w^{-1}\pm \sqrt{\left(\tau_m^{-1}+\tau_w^{-1}\right)^2 -4\left(1+g_w\right)\tau_m^{-1} \tau_w^{-1}}\right).
    \end{equation}
     and the eigenvectors
     \begin{equation}
     \xi^\pm = \left(\frac{g_w}{\tau_m},\frac{1}{2}\left(-\frac{1}{\tau_m}+\frac{1}{\tau_w} \mp \sqrt{\left(\frac{1}{\tau_m}-\frac{1}{\tau_w}\right)^2-4\frac{g_w}{\tau_m\tau_w}}\right)\right)^T. 
    \label{eigv_w}
     \end{equation}
   \noindent  The eigenvalues are complex if and only if $g_w > \left(4\tau_m \tau_w\right)^{-1} \left(\tau_w-\tau_m\right)^2$, and in that case their real part is $\frac{1}{2 \tau_m \tau_w} \left(\tau_m + \tau_w\right)$. As the adaptive time constant becomes slower, at a certain critical adaptation time constant both eigenvalues become real. We are interested in the behavior when the adaptation time constant is large. The absolute value of the inverse of the eigenvalues determines the time constants of the dynamics. Therefore, for large $\tau_w$ we can calculate the two real eigenvalues to first order of $\tau_w^{-1}$
    \begin{eqnarray}
        \lambda_w^+ =& - \frac{1+g_w}{\tau_w}   + O\left(\tau_w^{-2}\right)\\
    \lambda_w^- =& -\tau_m^{-1} +g_w \tau_w^{-1}+ O\left(\tau_w^{-2}\right).
        \end{eqnarray}
    
    In this limit of slow adaptation, the time constant of one eigenmode  is proportional to $\tau_w$, whereas the second mode scales with $\tau_m$. We are interested in the amplitude of each mode with respect to the other. 
    
   \noindent By explicitly calculating the first entry of the propagator matrix we obtain the adaptive filter in terms of the eigenvectors and eigenvalues, 
   \begin{equation}
   h_w\left(t\right) = \frac{1}{\xi_1^+\xi_2^- - \xi_1^- \xi_2^+ } \left(\xi_1^+ \xi_2^- e^{\lambda^+ t} - \xi_1^- \xi_2^+ e^{\lambda^- t} \right),\label{filt_w}
    \end{equation}
   \noindent where we use the notation $\xi_{1}^+$ to indicate the first component of the eigenvector associated to the eigenvalue $\lambda^+$. Approximating to leading order of $\tau_w^{-1}$ the eigenvectors in Eq~\eqref{eigv_w}, we obtain the eigenvectors
    
     \begin{align}\label{xi1}
        \xi_- &= \frac{1}{\tau_m}\left(g_w, 0\right)^T - \frac{1}{\tau_w} \left(0 , g_w\right)^T  = g_w\left(\frac{1}{\tau_m},-\frac{1}{\tau_w}\right)^T\\
        \xi_+ &= \frac{1}{\tau_m}\left(g_w,-1\right)^T + \frac{1}{\tau_w} \left(0, 1+g_w\right)^T  = \left(\frac{g_w}{\tau_m}, -\frac{1}{\tau_m}+ \frac{1+g_w}{\tau_w} \right)^T.\label{xi2}
        \end{align}

    \noindent  Then, using Eqs~\eqref{filt_w},~\eqref{xi1} and~\eqref{xi2}, we determine the linear filter:
        \begin{equation}
        h_w\left(t\right)  =  \frac{g_w }{\tau_m \left(2g_w+1\right)-\tau_w}e^{-\frac{1+g_w}{\tau_w}t}+\frac{1}{\tau_m}\frac{1-\left(1+g_w\right)\frac{\tau_m}{\tau_w}}{1-\left(1+2g_w\right)\frac{\tau_m}{\tau_w}} e^{-\left(\frac{1}{\tau_m}-\frac{g_w}{\tau_w}\right)t}.\label{adap_f}
        \end{equation}
      
    Interestingly, in contrast with synaptic filtering, the amplitude of the two modes are not equal. The amplitude of the slow mode (first term in Eq~\ref{adap_f}), whose timescale is proportional to $\tau_w$, decays proportionally to $\tau_w^{-1}$ with respect to the fast mode, when $\tau_w\ll \tau_m \left(2g_w+1\right)$. Therefore, the area under the linear filter corresponding to this mode is independent of $\tau_w$ for very large adaptation time constants:
    
    \begin{equation}
    \lim_{\tau_w \to \infty} \int_0^\infty h_w^{+} \left(t\right)\,dt  = \lim_{\tau_w \to \infty} \frac{g_w \tau_w}{\tau_m \left(g_w+1\right)\left(2g_w+1\right)-\left(g_w+1\right)\tau_w}     = -\frac{g_w }{g_w+1} .
    \end{equation}
    
    \noindent It follows that, if the adaptation timescale is increased, its relative contribution to the activity will decrease by the same factor, so that very slow adaptive currents will effectively be masked by the fast mode. 
    
\subsection*{Equilibrium activity}

The two systems possess a non-trivial equilibrium state at which the input current of all units stays constant. Since all units are statistically equivalent, the equilibrium activity is the same for all units. For synaptic filtering, the input current at equilibrium is given by a transcendental equation, that is obtained by setting to zero the left hand side of Eq~\eqref{Synap}:

\begin{equation}
x_0 = J\left(C_E - gC_I\right) \phi\left(x_0\right).\label{fp_syn}
\end{equation}

\noindent This equilibrium coincides with the fixed point of the system without synaptic filtering. 

For adaption, instead, from Eq~\eqref{Adap} we obtain that the equilibrium is determined by 
 
\begin{equation}
x_0 = \frac{1}{1+g_w} \left(J\left(C_E - gC_I\right)\phi\left(x_0\right) +g_w\gamma\right).\label{fp_adap}
\end{equation}

\noindent We further assume unless otherwise specified that the fixed point of the system is in the linear regime of the transfer function, so that $\phi\left(x\right) = x-\gamma$. In that case $x_0 = \left(J\left(C_E -gC_I\right)-g_w\right) \left(x_0-\gamma\right)$, so that larger adaptation coupling corresponds to weaker input currents, i.e. decreasing stationary firing rate. The adaptation time constant does not affect the fixed point.

\subsection*{Dynamics of homogeneous perturbations}

 We study the neuronal dynamics in response to a small perturbation uniform across the network
 
\begin{equation}
 x_i\left(t\right) = x_0 + \delta x\left(t\right).
 \end{equation} 
 
 \paragraph{Synaptic filtering} Linearizing Eq~\ref{Synap} we obtain 
 	\begin{equation}
 	\begin{cases}
 	&\tau_m \, \delta\dot{ x}_i \left(t\right)=-\delta x\left(t\right) +\delta s_i\left(t\right)\\
 	& \tau_s \delta\dot{s}_i\left(t\right) = -\delta s_i\left(t\right) + \phi_0^\prime \sum_j J_{ij} \delta x\left(t\right),
 	\end{cases} \label{lin_synap}
 	\end{equation}
 	
 	\noindent where we use the notation $\phi_0^\prime := \left. \frac{d\phi\left(x\right)}{dx} \right\rvert_{x_0}$. Because the perturbation $\delta x$ in Eq~\eqref{lin_synap}  is independent of $j$, using Eq~\eqref{eff_input} the dynamics for all units are equivalent to the population-averaged dynamics and are given by
 	\begin{equation}
 	 	\begin{cases}
 		&\tau_m \, \delta\dot{ x} \left(t\right)=-\delta x\left(t\right) +\delta s\left(t\right)\\
 		& \tau_s \delta\dot{s}\left(t\right) = -\delta s\left(t\right) + \phi_0^\prime  J\left(C_E-gC_I\right) \delta x.
 	\end{cases} \label{lin_synap2}
 \end{equation}
 	
 	\noindent From Eq~\eqref{lin_synap2} we can define the dynamic matrix

	\begin{equation}
A_s = \frac{1}{\tau_m}\left(\begin{array}{c c}
-1  & 1 \\
\phi_0^\prime J\left(C_E-gC_I\right)\frac{\tau_m}{\tau_s} & -\frac{\tau_m}{\tau_s}
\end{array}\right).
\end{equation}

 \noindent The only difference in the linearized dynamics of the population-averaged current with respect to the single neuron dynamics (Eq~\ref{dyn_syn}) is the non-diagonal entry $\phi_0^\prime J\left(C_E -gC_I\right)$. When either the derivative at the fixed point cancels, or when the total effective input is zero, the population dynamics equals the dynamics of a single neuron. The eigenvalues of the population-averaged dynamics are
 
 	\begin{equation}
 	\lambda_s^\pm = -\frac{\tau_m+\tau_s}{2\tau_s \tau_m} \pm \sqrt{\left(\frac{\tau_m-\tau_s}{2\tau_s \tau_m}\right)^2+\frac{J\left(C_E-gC_I\right)}{\tau_m \tau_s}}.
 	\end{equation}
 	
 	\noindent and the eigenvectors
 	
 	\begin{equation}
 	\xi_s^\pm = \left(-1, \frac{\tau_m-\tau_s}{2\tau_s \tau_m} \mp \sqrt{\left(\frac{\tau_m-\tau_s}{2\tau_s \tau_m}\right)^2+\frac{J\left(C_E-gC_I\right)}{\tau_m \tau_s}}\right)^T.
 	\end{equation}
 For very large synaptic time constants, the eigenvalues are approximated to leading order as
 
 \begin{align}
 \lambda_s^+ &= \frac{J\left(C_E-gC_I\right)-1}{\tau_s} + O\left(\tau_s^{-2}\right) \\
 \lambda_s^- &= -\frac{1}{\tau_m} - \frac{J\left(C_E-gC_I\right)}{\tau_s} 
 \end{align}
 
 \noindent Approximating as well the eigenvectors to leading order, we obtain 
 \begin{align}
 \xi^+ &= \left(\frac{1}{\tau_m}, \frac{1}{\tau_m}-\frac{1-J\left(C_E-gC_I\right)}{\tau_s}\right)^T\\
 \xi^- &= \left(\frac{1}{\tau_m}, -\frac{J\left(C_E-gC_I\right)}{\tau_s}\right)^T
 \end{align}
 the filter of the linear response to weak homogeneous perturbations reads:
 
 \begin{align}
 h_s\left(t\right) &= \frac{1}{\tau_s} \frac{\xi_1^-\xi_1^+}{\xi_1^+ \xi_2^- - \xi_1^- \xi_2^+} \left(e^{\lambda^- t}-e^{\lambda^+ t}\right)\\
 &= \frac{1}{\tau_s}\frac{\tau_s-\tau_m\left(1-J\left(C_E-gC_I\right)\right)}{\tau_s-\tau_m\left(1-2J\left(C_E-gC_I\right)\right)} \left(e^{\lambda^- t}-e^{\lambda^+ t}\right)
 \end{align}
 
 \noindent Note that the amplitude of the two exponential terms is the same, independently of the effective coupling and time constants.
  
\paragraph{Adaptation} For the system with adaptive neurons, the linearized system reads 

 	\begin{equation}
 	\begin{cases}
 	&\tau_m \, \delta\dot{x}_i \left(t\right)=-\delta x_i\left(t\right) -g_w \delta w_i\left(t\right)\left(t\right) +\phi_0^\prime \sum_j J_{ij} \delta x\left(t\right) \\
 	& \tau_w \delta\dot{w}_i\left(t\right) = -\delta w_i\left(t\right) + \delta x\left(t\right).
 	\end{cases} \label{lin_adap}
 	\end{equation}

\noindent As for the network with synaptic filtering, the dynamics of the perturbation are equivalent for each unit, so that we can write down the dynamic matrix for the population-averaged response to homogeneous perturbations
 	
 \begin{eqnarray}
A_w = \frac{1}{\tau_m}\left(\begin{array}{c c}
-1+\phi_0^\prime J\left(C_E-gC_I\right)  & -g_w  \\
\frac{\tau_m}{\tau_w} & -\frac{\tau_m}{\tau_w}
\end{array}\right).
\end{eqnarray}

\noindent The difference with respect to the linear single neuron dynamics (Eq~\ref{lin_adap}) is that the effective recurrent coupling appears now in the first diagonal entry of the dynamic matrix.

  When the fixed point is located within the linear range of the transfer function, the derivative is one, so that we do not further specify the factor $\phi_0^\prime$ in the following equations. Consequently, the dynamics of the system to small perturbations do not depend on the exact value of the fixed point, which does not hold for more general transfer functions.

The eigenvalues of the system read
\begin{equation}
\lambda_w^\pm =\left( -\frac{1-J_{\text{eff}}}{2\tau_m}-\frac{1}{2\tau_w}\right)\left(1  \pm \sqrt{1 + \frac{4\tau_m \left(J_{\text{eff}} -1 -g_w\right)}{\tau_w \left(J_{\text{eff}}-1-\frac{\tau_m}{\tau_w}\right)^2}}\right),
\end{equation}

\noindent with eigenvectors

\begin{equation}
\xi_w^\pm =\left(2g_w, \frac{\tau_m}{\tau_w}+J_{\text{eff}}-1 \mp \sqrt{\left(\frac{\tau_m}{\tau_w}-J_{\text{eff}}+1\right)^2-4\frac{\tau_m}{\tau_w} \left(g_w-J_{\text{eff}}+1\right)}\right)^T
\end{equation}

In the limit of very slow adaptation, given that the two eigenvalues are real,  they can be approximated to leading order as

\begin{align}
\lambda_w^+ &= 1+\frac{\tau_m}{\tau_w \left(J\left(C_E-gC_I\right)-1\right)}+O\left(\tau_w^{-2}\right)\\
\lambda_w^- &= -\frac{1}{\tau_w} \left(1-\frac{g_w}{J\left(C_E-gC_I\right) -1}\right)+O\left(\tau_w^{-2}\right)
\end{align}

\noindent and the corresponding eigenvectors read

\begin{align}
\xi_w^+ &= \left(1, \frac{1}{J_\text{eff}-1} \frac{\tau_m}{\tau_w}\right)^T\\
\xi_w^- &= \left(g_w, J_\text{eff}-1 + \frac{\tau_m}{\tau_w} \left(1-\frac{g_w}{J_\text{eff}-1}\right) \right)^T.
\end{align}

\noindent Therefore, if the perturbation is stable (see next section) we can write down the corresponding linear filter as

        \begin{equation}
h_w\left(t\right)  = \frac{1}{\tau_m}\frac{J_\text{eff}-1+\frac{\tau_m}{\tau_w}\left(1-\frac{g_w}{J_\text{eff}}\right)}{J_\text{eff}-1+\frac{\tau_m}{\tau_w} \left(1-\frac{2g_w}{J_\text{eff}}\right)}e^{\lambda_w^+ t}-\frac{g_w}{\tau_w \left(J_\text{eff}-1\right)^2 + \tau_m \left(J_\text{eff}-1-2g_w\right)  }e^{\lambda_w^- t}.\label{adap_fJ}
\end{equation}

The area under the slow mode is again independent of the adaptation time constant in this limit,

  \begin{equation}
\lim_{\tau_w \to \infty} \int_0^\infty h_w^{-} \left(t\right)\,dt  = -\frac{g_w }{\left(J_\text{eff}-1\right)\left(J_\text{eff}-1-g_w\right)}.
\end{equation}

%
\subsection*{Stability of homogeneous perturbations}
The equilibrium point is stable when the real part of all eigenvalues is negative. Equivalently, in a two dimensional system --as it is the case for the population-averaged dynamics--, the dynamics are stable when the trace of the dynamic matrix is negative and the determinant positive. 

\paragraph{Synaptic filtering} In the system with synaptic filtering, the trace and determinant are

\begin{align}
\text{Tr}_s  &= -\frac{1}{\tau_m} -\frac{1}{\tau_s} \\
\text{Det}_s &= \frac{1-J\left(C_E-gC_I\right)}{\tau_m \tau_s}.
\end{align}

\noindent The trace is therefore always negative. The determinant is positive, and therefore the population-averaged dynamics are stable, when the effective coupling $J\left(C_E -gC_I\right)$ is smaller than unity. In contrast, if the effective coupling is larger than unity, i.e. if positive feedback is too strong, the equilibrium firing rate is unstable, so that any small perturbation to the equilibrium firing rate will lead the system to a different state. Right at the critical effective coupling, one eigenvalues is zero and the other one equals $\text{Tr}_s$, implying that the population-averaged dynamics undergo a saddle-node bifurcation. 
Beyond the bifurcation, the network reaches a state where the firing rates of all neurons saturate. 

\paragraph{Adaptation} In the adaptive population dynamics, the recurrent connectivity has a different effect on the stability of the adaptive population dynamics. The trace and determinant of the dynamic matrix are
\begin{eqnarray}
\text{Tr}~_w &=& -\frac{1}{\tau_m}-\frac{1}{\tau_w}+\tau_m^{-1} J\left(C_E-gC_I\right),\\
\text{Det}~_w &=& \left(\tau_m\tau_w\right)^{-1} \left(1-J\left(C_E-gC_I\right)+g_w\right).
\end{eqnarray}

\noindent Both the timescale $\tau_w$ and the strength $g_w$ of adaptation affect the trace and determinant of the dynamic matrix, and therefore the stability. The system is unstable if the determinant is negative (one positive and one negative real eigenvalue) or if the determinant is positive and the trace is positive. The determinant is negative, and therefore the system becomes unstable through a saddle-node bifurcation, when $J\left(C_E -gC_I\right) > 1+g_w$. Note that the adaptation strength increases the stability of the system: a stronger positive feedback loop is required to destabilize the fixed point, in comparison to the network with synaptic filtering. The determinant and trace are positive if $J\left(C_E -gC_I\right) < 1+g_w$ but $J\left(C_E -gC_I\right) > 1+\frac{\tau_m}{\tau_w}$, respectively, leading to a Hopf bifurcation: the system produces sustained marginal oscillations at the bifurcation in response to small perturbations around the fixed point. Beyond the Hopf bifurcation, the oscillations are maintained in time, unless the system shows a fixed point when all neurons saturate ($x_0 = \frac{1}{1-g_w} \left( J\left(C_E-gC_I\right) \phi_{max}+g_w\gamma \right)$). This fixed point exists if $x_0>\phi_{max} + \gamma$.

\subsection*{Heterogeneous activity}

We next study the network dynamics beyond the population-averaged activity, along modes where different units have different amplitudes. We study perturbations of the type

\begin{equation}
x_i\left(t\right) = x_0 + \delta x_i \left(t\right).
\end{equation}

\noindent We define the $2N$-dimensional vector $\textbf{x} = \left(\delta x_1, ..., \delta x_N^1, \delta w_1^1, ..., \delta w_N^1\right)^T$. Since the dynamics of each unit is now different, the dynamic matrix of the linearized system, $A$, is described by a squared matrix of dimensionality $2N$. Therefore, the perturbations generate dynamics along $2N$ different modes whose timescales are determined by the eigenvalues of the matrix $A$. The eigenvalues are determined by the characteristic equation $\left|A-\lambda I\right|=0$. In order to calculate these eigenvalues, we make use of the following identity which holds for any block matrix $Z=A-\lambda I$, that is composed by the four square matrices \textbf{P},\textbf{Q}, \textbf{R}, and \textbf{S} and the block \textbf{S} is invertible:
\begin{equation}
\left\lvert Z \right\rvert := \left\lvert \left(\begin{array}{c c}
\textbf{P}  & \textbf{Q} \\
\textbf{R} & \textbf{S}
\end{array}\right) \right\rvert = \left\lvert \textbf{S} \right\rvert \left\lvert \textbf{P}-\textbf{Q}\textbf{S}^{-1}\textbf{R} \right\rvert.\label{trick1}
\end{equation}

\noindent Consequently, if we set Eq~\eqref{trick1} to zero, since we assumed that $\left\lvert S\right\rvert \neq 0$, we obtain

\begin{equation}
\left\lvert Z \right\rvert =0 \implies \left\lvert \textbf{P}-\textbf{Q}\textbf{S}^{-1}\textbf{R} \right\rvert =0. \label{trick} 
\end{equation}
\noindent The identity in Eq~\eqref{trick1} can be shown by using the decomposition

\begin{equation}
Z = \left(\begin{array}{c c}
\textbf{I}  & 0 \\
0 & \textbf{S}
\end{array}\right) 
\left(\begin{array}{c c}
\textbf{I}  & \textbf{Q} \\
0 & \textbf{I}
\end{array}\right)
\left(\begin{array}{c c}
\textbf{P}-\textbf{Q}\textbf{S}^{-1}\textbf{R}  & 0 \\
\textbf{S}^{-1} \textbf{R} & \textbf{I}
\end{array}\right),
\end{equation}

\noindent together with the fact that when a non-diagonal block is zero. The determinant of such a matrix is the product of determinants of the diagonal blocks.

\paragraph{Synaptic filtering}
 The dynamical matrix for the network with synaptic filtering, obtained by linearizing Eqs~\eqref{Synap}, is 
\begin{equation}
A_s = \frac{1}{\tau_m}\left(
\begin{array}{c|c}
 - \textbf{I}  & \textbf{I} \\
\hline
\phi_0^\prime \textbf{J} \frac{\tau_m}{\tau_s} & -\frac{\tau_m}{\tau_s}\textbf{I}
\end{array}
\right),\label{linfull_syn}
\end{equation}

\noindent The matrix $\textbf{J}$ is the connectivity matrix. Again, we assume in the following that the fixed point is located in the linear range of the transfer function, so that $\phi_0^\prime = 1$. 

The characteristic equation, obtained by combining Eqs~\eqref{trick} and \eqref{linfull_syn}, reads

\begin{eqnarray}
\left| -\left(1+\tau_m \lambda_s \right)\textbf{I} + \left(\frac{\tau_m}{\tau_s}+\tau_m \lambda_s\right)^{-1}\frac{\tau_m}{\tau_s} \textbf{J}\right| = -\left(1+\tau_m \lambda_s\right) + \frac{\lambda_J}{1+\tau_s \lambda_s} = 0, \label{step2}
\end{eqnarray}

\noindent where $\lambda_J$ are the eigenvalues of the connectivity matrix. Solving for $\lambda_J$ we obtain the equation which maps the eigenvalues of the synaptic filtering network dynamics $\lambda_s$ onto the eigenvalues of the connectivity matrix $\lambda_J$,

\begin{equation}
\lambda_J = \left(1+\tau_m \lambda_s\right) \left(1 + \tau_s \lambda_s \right).\label{invmap_syn}
\end{equation}
\noindent In contrast, solving for the eigenvalues of the dynamic matrix $\lambda_s$ we obtain the inverse mapping

\begin{equation}
\lambda_s^2 + \frac{\tau_s+\tau_m}{
\tau_s \tau_m} \lambda_s + \frac{1-\lambda_J}{ \tau_s \tau_m} =0.\label{map_syn}
\end{equation}

\noindent In other words, Eqs~\ref{map_syn} and ~\ref{invmap_syn} constitute two different approaches to assessing the stability of the system \cite{Bimbard2016}. One approach is to examine whether the domain of eigenvalues $\lambda_s$ resulting from Eq~\eqref{map_syn} intersect the line $\text{Re}\left(\lambda_s\right)=0$ (Fig~\ref{fig3}, insets in B). The eigenvalues $\lambda_J$ of the connectivity matrix  are distributed within a circle in the complex plane, whose radius is proportional to the synaptic strength, $\lambda_J < J\sqrt{C_E + g^2 C_I}$ plus an outlier real eigenvalue at $J\left(C_E-gC_I\right)$ that corresponds to the homogeneous perturbations studied above (see \cite{Rajan2006}). We focus in this section on the bulk of eigenvalues that corresponds to modes of activity with different amplitudes for different units. We can therefore parametrize the eigenvalues $\lambda_J$ as

\begin{equation}
\lambda_J\left(\theta\right) = J\sqrt{C_E+g^2C_I} e^{i\theta} \label{param_dir}
\end{equation}

and introduce the parametrization into Eq~\eqref{map_syn} to obtain an explicit expression for the curve that encloses the eigenspectrum $\lambda_s$. Note that in an abuse of notation, we denote the limits of the eigenspectrum as $\lambda$ and not the eigenvalues themselves that constitute the eigenspectrum. 

\begin{figure}[!h]
	\begin{adjustwidth}{-2.1in}{0in}
		\includegraphics[width=\linewidth]{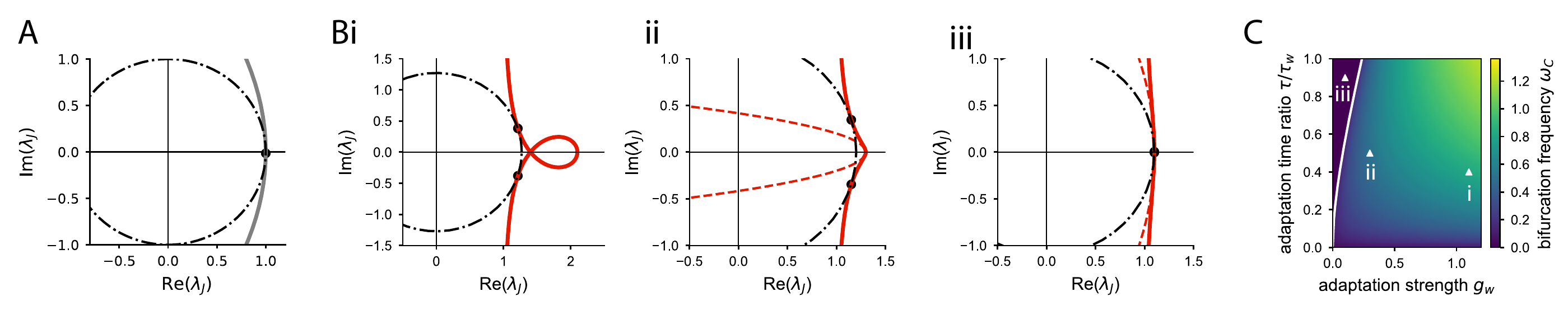}
		\caption{{\bf Geometrical description of the bifurcation of the heterogeneous activity. } A: Instability bound for the system with synaptic filtering (grey line, Eq.\ref{invsyn1}) and eigenspectrum for the weakest unstable synaptic coupling $J$. For any parameter combination, the instability bound, a parabola, is first touched by the growing circle of eigenvalues at $\omega=1$ and value $J\sqrt{C_E+g^2C_I} = 1.$ B: Three different configurations of the instability bound for the system with adaptation in the complex plane of eigenvalues of the connectivity matrix, $\lambda_J$. The black dots indicate the intersection between the instability boundary (full red line) and the eigenspectrum of $\lambda_J$ (dashed black line) with weakest coupling that is unstable. (i) The instability boundary intersects the real axis twice, leading to a Hopf bifurcation. (ii) It intersects the real axis just once and still leads to a Hopf bifurcation, because the intersection with the real axis is not the closest point of the curve to the origin. (iii) It intersects the real axis once and leads to a zero-frequency bifurcation, because the crossing of the real axis is the closest point to the origin. In (ii) and (iii) we draw the parabolic approximation of the instability bound (red dashed line, Eq~\ref{parab_aproxadap}). If the curvature of this parabola is exterior to the $\lambda_J$ eigenspectrum, as in (iii), the system undergoes a zero-frequency bifurcation.   C: Oscillatory frequency at which the network with adaptation undergoes a bifurcation. To the right of the white line (Eq~\ref{sadHopf}), the network displays a Hopf bifurcation, whereas to the left, the bifurcation happens at zero-frequency.  The triangles indicate the parameter combinations used in B. }
		\label{figSI1}
	\end{adjustwidth}
\end{figure}

The alternative approach is to use the inverse mapping from the eigenvalues $\lambda_s$ to the eigenvalues of the connectivity $\lambda_J$, by mapping the line $\text{Re}\left(\lambda_s\right)=0$ into the space of eigenvalues $\lambda_J$ (Fig~\ref{figSI1}). More specifically, the line $\text{Re}\left(\lambda_s\right)=0$ can be parametrized as 
\begin{equation}
\lambda_s = \pm i \omega,\label{param_inv}
\end{equation}
\noindent and introduced into Eq~\eqref{invmap_syn}.  In this case, the stability is assessed by whether the eigenspectrum of the connectivity matrix $\textbf{J}$ crosses the stability boundary or not (insets Fig~\ref{figSI1}). This alternative approach is useful for some calculations due to the simple geometry of the connectivity eigenspectrum $\lambda_J$.
 
 Taking the alternative approach, introducing Eq~\eqref{param_inv} into Eq~\eqref{invmap_syn}, we obtain the stability bound in the complex plane of eigenvalues $\lambda_J$:

\begin{equation}
\lambda_J^{sb} = \left(1+i\tau_m \omega\right) \left(1+i \tau_s \omega\right).\label{invsyn1}
\end{equation} 

The first point of the stability curve $\lambda_J^{sb}\left(\omega\right)$ intersecting with a circle of increasing radius centered at the origin is the closest point of the curve to the origin, i.e. the minimum of $\left|\lambda_J^{sb}\right|^2$ with respect to $\omega$. The squared distance to the origin is
\begin{equation}
\left\lvert\lambda_J^{sb}\right\rvert^2 = \left(1+\tau_m^2 \omega^2\right)\left(1+\tau_w^2 \omega^2\right),
\end{equation}
whose minimum happens trivially at $\omega=0$, $\lambda_J = 1$ (Fig~\ref{figSI1}A). In conclusion, the system is unstable if
\begin{equation}
J\left(C_E+g^2 C_I\right)>1.
\end{equation}

\noindent Note that this is the same condition as in the case without synaptic filtering, The synaptic filtering system approaches the no-filtering system when $\tau_s \to 0$. Although we are considering in this work synaptic timescales that are larger than the membrane time constant, the analysis is valid for arbitrarily fast synaptic time constants. In that limit, the stability curve in Eq~\eqref{invsyn1} approaches the curve $\lambda_J^{sb} = 1$, retrieving the stability boundary found in \cite{Sompolinsky1988}.
	 
To study the limit of slow synaptic time constant,  $\tau_s \gg \tau_m$, we analyze the direct approach, i.e. study how the parameters of adaptation modify the eigenspectrum of the dynamic matrix $A_s$ in the complex plane of eigenvalues $\lambda_s$. To this end, we introduce the parametrized connectivity eigenspectrum (Eq~\ref{param_dir}) into Eq~\eqref{map_syn}, and approximate it to leading order of $\frac{\tau_m}{\tau_s}$. We obtain that the eigenspectrum of eigenvalues $\lambda_s$ are enclosed by the curves
 
\begin{align}
\lambda_s^+ &\approx \frac{1}{\tau_s} \left(J\sqrt{C_E+g^2C_I}e^{i\theta}-1\right)\\
\lambda_s^- &\approx  -\frac{1}{\tau_m} - \frac{1}{\tau_s} \left(J\sqrt{C_E+g^2C_I}e^{i\theta}-1\right).\label{approx2circ}
\end{align}

\noindent The equations above approximate the full eigenspectrum by two disjoint circles of radius $\tau_s^{-1} J\sqrt{C_E+g^2C_I}$, the one corresponding to the $\lambda_s^+$ eigenvalues centered at $-\frac{1}{\tau_s}$, and the other circle $\lambda^-$ centered at $-\frac{1}{\tau_m}+\frac{1}{\tau_s}$. The circle centered closer to the instability bound, $\lambda_s^+$ sets the slow timescales of the network, and its associated timescale is proportional to $\tau_s$. This gives an intuitive explanation to why the network timescale scales linearly with the synaptic time constant (Fig \ref{fig5}). 

\paragraph{Adaptation}

For adaptation, we repeat the same procedure as for the synaptic filtering to determine the stability to heterogeneous perturbations. The dynamical matrix reads

\begin{equation}
A_w = \frac{1}{\tau_m}\left(
\begin{array}{c|c}
\phi_0^\prime \textbf{J} - \textbf{I}  & -g_w \textbf{I} \\
\hline
\phi_0^\prime \frac{\tau_m}{\tau_w} \textbf{I} & -\frac{\tau_m}{\tau_w}\textbf{I}
\end{array}
\right),\label{linfull_adap}
\end{equation}

Using Eqs~\eqref{trick} and \eqref{linfull_adap} we can obtain the characteristic equation. Solving for $\lambda_J$ we obtain the mapping between the $\lambda_w$ eigenvalues and the connectivity eigenvalues

\begin{equation}\label{ref_inv}
\lambda_J =  1 +\tau_m \lambda_w + g_w \frac{\tau_m}{\tau_w} \left(\tau_m \lambda_w + \frac{\tau_m}{\tau_w}\right)^{-1},
\end{equation}

\noindent while solving for $\lambda_w$ we obtain the expression for the inverse mapping:

\begin{equation}\label{ref_lm}
\left(\tau_m \lambda_w\right)^2 + \left(1+\frac{\tau_m}{\tau_w}-\lambda_J\right)\tau_m \lambda_w + \frac{\tau_m}{\tau_w}\left(1+g_w  - \lambda_J\right) = 0.
\end{equation} 

We first explore the inverse mapping. Inserting the parametrization in Eq~\eqref{param_inv} into Eq~\eqref{ref_inv}, the stability curve in the complex plane of connectivity eigenvalues reads 
 
 \begin{equation}
 \lambda_J^{sb} \left(\omega\right) = 1+ \frac{g_w}{1-\tau_w^2 \omega^2} + i\omega \left(\tau_m - \tau_w \frac{g_w}{1-\tau_w^2 \omega^2}\right).\label{inv_adap}
 \end{equation}
 
 The bifurcation parameters can then be found by determining the closest point of the stability boundary to the origin. The corresponding value of $\omega$ determines the oscillatory frequency of the first unstable mode. This value can be zero, corresponding to a zero-frequency bifurcation, which generally leads to slowly fluctuating activity referred to as rate chaos (\cite{Sompolinsky1988}, \cite{Kadmon2015}, \cite{Harish2015}, \cite{Mastrogiuseppe2017}). Alternatively, when the parameter $\omega$ that minimizes the norm of $\lambda_J^{sb}$ is non-zero, the system undergoes a Hopf bifurcation. 
  
 It is useful to consider the different geometries of the stability curve in Eq~\eqref{inv_adap} in order to identify the closest point of the curve to the origin. Note that the curve shows symmetry with respect to the real axis, $\lambda_J^{sb}\left(-\omega\right) = \lambda_J^{sb*}\left(\omega\right)$. 
 
 The curve might cross the real axis $\text{Im}\left(\lambda_J\right)=0$ either in one or two different values of $\left|\omega\right|$. Solving $\text{Re}\left(\lambda_J^{sb}\right) = 0$, we find that the curve crosses twice the real axis, when $\tau_m < \tau_w g_w$  (Fig~\ref{figSI1} Bi). In that case, one crossing is the point $\tau_m \lambda_J = 1+\frac{\tau_m}{\tau_w}$ and the other $\tau_m \lambda_J = 1 + g_w$. This second intersection corresponds to $\omega = 0$. Therefore, it is clear that, since the first crossing of the real axis is closer to the origin than the point at $\omega=0$, the bifurcation necessarily occurs at non-zero frequency for $\tau_m < \tau_w g_w$.
 
When the curve crosses only once the zero axis, the point $\lambda_J = 1+g_w$, corresponding to a zero-frequency, is not necessarily the closest one to the origin (Fig~\ref{figSI1} Bii). One approach to determine analytically whether the system undergoes a Hopf or a zero-frequency bifurcation is to look at the curvature at the point $\omega=0$ and compare it to the curvature of a circle with radius $1+g_w$. To do so, we approximate both the stability line and the circle by a parabola, and compare their curvatures (dashed curve, Fig\ref{figSI1} Bii and Biii). First, we write the stability boundary in its implicit form, $\lambda_J^{sb} : = x_J^{sb} + i y_{J}^{sb}$, as
 
    \begin{equation}
   \left(y^{sb}\right)^2 - \left(x^{sb}-1-g_w\right) \left(x^{sb}-1-\frac{\tau_m}{\tau_w}\right)^2 \frac{1}{x^{sb}-1}=0.
   \end{equation}
  \noindent Then, we consider small deviations of the coordinates $x^{sb} = 1+g_w + \epsilon_x$ and $y^{sb} = \epsilon_y$. If we approximate up to first order of $\epsilon_x$ and second order of $\epsilon_y$ we obtain the parabola
   
   \begin{equation}
   \epsilon_y^2 = \frac{\left(g_w-\frac{\tau_m}{\tau_w}\right)^2}{g_w} \epsilon_x +O\left(\epsilon_x^2\right).\label{parab_aproxadap}
   \end{equation}
     
Repeating the same procedure for the circle of eigenvalues, with radius $r= 1+g_w$ we obtain $\epsilon_y^2 = 2\left(1+g_w\right) \epsilon_x + O\left(\epsilon_x^2\right)$. By requiring the circle of eigenvalues to be interior to the boundary curve (for the same $\epsilon_x$, $\epsilon_{y, \text{circle}}^2 < \epsilon_{y, \text{sb}}^2$), we obtain that the instability parabola is exterior to the circle, therefore the system undergoes a zero-frequency bifurcation (Fig~\ref{figSI1}C), when 
   
   \begin{equation}
   \frac{\left(g_w-\frac{\tau_m}{\tau_w}\right)^2}{g_w} \epsilon_x  < 2\left(1+g_w\right) \epsilon_x
   \end{equation}
  
  \noindent which simplifies to
  
  \begin{equation}
  \frac{\tau_m}{\tau_w} > g_w + \sqrt{2g_w\left(g_w+1\right)}. \label{sadHopf}
  \end{equation}
  
  In the limit of the adaptation timescale approaching the membrane time constant, the left side of the inequality above approaches one. Introducing this value in Eq~\ref{sadHopf}, we find that for adaptive couplings stronger than $g_w>\sqrt{5}-2$ only a Hopf bifurcation is possible.

\subsection*{Dynamical Mean Field Theory}

The linearization of the dynamical system from the previous section is only valid up to the instability boundary. A commonly used method to study the dynamics that arise beyond the bifurcation is dynamical mean field theory (DMFT) \cite{Sompolinsky1988, Rajan2010, Stern2014,  Harish2015, Aljadeff2015, Kadmon2015,  Mastrogiuseppe2017}. DMFT approximates the deterministic input to each element of the system by a Gaussian stochastic process, whose first and second moment are determined self-consistently. 

The dynamics of the $i$-th neuron in the synaptic and adaptive network are approximated as

\begin{equation}
\begin{cases}
&\tau_m \, \dot{x}_i \left(t\right)=-x_i\left(t\right) +s_i\left(t\right)\\
& \tau_s \dot{s}_i\left(t\right) = -s_i\left(t\right) +\xi_{i} \left(t\right),
\end{cases} \label{DMFT_syn}
\end{equation}

\begin{equation}
\begin{cases}
&\tau_m \, \dot{x}_i \left(t\right)=-x_i\left(t\right) -g_w w_i\left(t\right)+\xi_{i}\left(t\right)\\
& \tau_w \dot{w}_i\left(t\right) = -w_i\left(t\right) + x_i\left(t\right) - \gamma,
\end{cases} \label{DMFT_adap}
\end{equation}

\noindent where $\xi_{i}\left(t\right)$ is a Gaussian variable. In the thermodynamic limit, the noise sources are independent between neurons, so that for $i\ne j$ $\left[\xi_i\left(t\right) \xi_j \left(t^\prime\right)\right] = 0 $.

The next step is to determine the self-consistent equations, that links the distribution of $\xi_i$ to the statistics of the original system in Eqs~\eqref{Adap} and \eqref{Synap}. First, we relate the statistics of the noise, currents $x_i$ and rates $\phi\left(x_i\right)$ based on the dynamics. Then, we close the equations by explicitly assuring that the transfer function relates the currents and the rates.

To determine the first moment of the noise, we apply that $\xi_i\left(t\right) = \sum_j J_{ij} \phi\left(x_j\left(t\right)\right)$ and average over the population, as in \cite{Mastrogiuseppe2017}. The first moment of the noise then obeys

\begin{eqnarray}
\left[\xi_i\right] = \left\langle \sum_{j=1}^N J_{ij} \phi_j\left(t\right) \right\rangle = J\left(C_E-gC_I\right)\left\langle\phi\right\rangle.\label{selfcon1}
\end{eqnarray}

We calculate next the relation for the second moment of the noise, which again is the same as in \cite{Mastrogiuseppe2017}:

\begin{eqnarray}
\left[\xi_i\left(t\right)\xi_j\left(t+\tau\right)\right] = \left\langle \sum_{k=1}^N J_{ik} \phi_k \left(t\right) \sum_{l=1}^N  J_{jl} \phi_l \left(t\right) \right\rangle = \delta_{ij} J^2\left(C_E+g^2C_I\right) \left(C\left(\tau\right)-\left\langle \phi \right\rangle^2\right),\label{selfcon2}
\end{eqnarray}

\noindent where $C\left(\tau\right) = \left\langle \phi_i\left(t\right) \phi_i\left(t+\tau \right)\right\rangle$ .

These equations show that the first and second moment of the Gaussian sources do not depend on the identity of neuron $i$, so that all neurons are statistically equivalent. Thus, we can reduce the full $2N$-deterministic system to a two-variable stochastic system, describing a prototypical neuron in the network.

The equations \eqref{selfcon1} and \eqref{selfcon2} describe how the noise is related to the properties of the connectivity and the statistics of the rates $\phi\left(x\right)$. The next step is to calculate how the first and second moment of the noise are related to the statistics of the input current, which we write as $\mu:= \left[x_i\right]$ for the first moment and $\Delta\left(\tau\right):=\left[x_i\left(t\right)x_i\left(t+\tau\right)\right] - \mu^2$ for the second moment.

For the mean of the input current, averaging over units Eqs~\eqref{DMFT_syn} and \eqref{DMFT_adap} and introducing the result in \eqref{selfcon1} for the synaptic and adaptive system respectively, we obtain

\begin{align}\label{1syn}
\mu_s &= \left[\xi\right] = J\left(C_E-gC_I\right) \left\langle \phi\right\rangle,\\
\mu_w &= \frac{1}{1+g_w} \left(g_w \gamma + \left[\xi\right]\right) = \frac{1}{1+g_w} \left(g_w\gamma+ J\left(C_E-gC_I\right)\left\langle \phi\right\rangle\right).\label{1dap}
\end{align}

By differentiating twice $\Delta\left(\tau\right)$ with respect to the lag $\tau$ and using Eqs~\eqref{DMFT_syn} and \eqref{selfcon2}, as in \cite{Sompolinsky1988, Mastrogiuseppe2017} we obtain:
\begin{equation}\label{2syn}
\ddot{\Delta}_s\left(\tau\right) = \Delta_s\left(\tau\right) + \left( Q_s \ast \Delta_s \right) \left(\tau\right)- J^2\left(C_E+g^2 C_I\right)\left(C\left(\tau\right)-\left\langle \phi \right\rangle^2 \right) ,
\end{equation}

\noindent where $Q_s\left(\tau\right):=\int_{-\infty}^{+\infty} dt h_s\left(t\right)h_s\left(t+\tau\right)$  is the autocorrelation function of the single neuron filter $h_s$ (Eq.\ref{synfilt}). Equivalently, for the adaptive system, using Eq~\eqref{DMFT_adap} and \eqref{selfcon2} we obtain

\begin{equation}\label{2adap}
\ddot{\Delta}_w\left(\tau\right) = \Delta_w\left(\tau\right) + \left(g_w\left(g_w Q_w + h_w^{sym} + \dot{h}_w^{sym}\right) \ast \Delta_w\right)\left(\tau\right)  - J^2\left(C_E+g^2 C_I\right)\left(C\left(\tau\right)-\left\langle \phi \right\rangle^2 \right). \end{equation}
\noindent where we define in relation to Eq~\eqref{adap_filt} $h_w^{sym}\left(\tau\right) = h_w\left(\left|\tau\right|\right)$, and the autocorrelation function of the adaptive filter $Q_w:=\int_{-\infty}^{+\infty} dt h_w\left(t\right) h_w\left(t+\tau\right)$.

Secondly, in order to close the self-consistent description, we can link the statistics of the rates $\phi_i\left(t\right)$ with the statistics of the currents $x_i\left(t\right)$ by writing the input currents explicitly as Gaussian variables. We can write down the input current at time $t$ and $t+\tau$ explicitly as (see \cite{Rajan2010}):

\begin{align}\label{Gausdef1}
x\left(t\right) &= \mu + \sqrt{\Delta\left(0\right) -\left\lvert \Delta\left(\tau\right)\right\rvert} z_1 + \text{sgn}\left(\Delta\left(\tau\right)\right)\sqrt{\left\lvert \Delta\left(\tau\right)\right\rvert} z_3\\
x\left(t+\tau\right) &= \mu + \sqrt{\Delta\left(0\right) -\left\lvert \Delta\left(\tau\right)\right\rvert} z_2 + \sqrt{\left\lvert \Delta\left(\tau\right)\right\rvert} z_3.\label{Gausdef2}
\end{align}

\noindent This explicit construction in terms of Gaussian variables $z_1$, $z_2$ and $z_3$ realizes the constraints $\left[x^2\left(t\right)\right] - \mu^2 = \Delta\left(0\right)$, $\left[x^2\left(t+\tau\right)\right] - \mu^2 = \Delta\left(0\right)$ and $\left[x\left(t\right) x\left(t+\tau\right)\right] - \mu^2 = \Delta\left(\tau\right)$. 
Now, explicitly calculating the first moment of the rates by replacing the average for a Gaussian integral and using Eq~\eqref{Gausdef1} we obtain

\begin{equation}\label{self1mom}
\left\langle \phi\right\rangle = \int Dz \phi\left(\mu + \sqrt{\Delta\left(0\right)} z \right)
\end{equation}
\noindent where we use the short-hand notation $\int Dz = \int_{-\infty}^{+\infty} \frac{e^{-\frac{z^2}{2}}}{2} dz$. 

For the second moment, introducing Eqs~\eqref{Gausdef1} and \eqref{Gausdef2} into the definition of autocorrelation function of the rate, we get

\begin{align}
C\left(\tau\right) = &\int Dz_3 \int Dz_1 \phi\left(\sqrt{\Delta\left(0\right) - \left\lvert\Delta\left(\tau\right) \right\rvert} z_1+ \text{sgn}\left(\Delta\left(\tau\right)\right)\sqrt{\left\lvert \Delta\left(\tau\right)\right\rvert} z_3\right) \nonumber \\
&\int Dz_2 \phi\left(\sqrt{\Delta\left(0\right) - \left\lvert\Delta\left(\tau\right) \right\rvert} z_2+ \sqrt{\left\lvert \Delta\left(\tau\right)\right\rvert} z_3\right). \label{self2mom}
\end{align}

Therefore, in order to determine the self-consistent solution, we need to find a mean and autocorrelation function for the currents that satisfy both Eqs~\eqref{self1mom} and ~\eqref{self2mom} and Eqs~\eqref{1syn}  and \eqref{2syn} (for the synaptic system) and Eqs~\eqref{1dap} and \eqref{2adap} (for the adaptive system). Once the statistics of the currents and rates are known, it is straight-forward to obtain the statistics of the noise, using Eqs~\eqref{selfcon1} and \eqref{selfcon2}.

In previous works \cite{Sompolinsky1988,  Kadmon2015, Harish2015, Schuecker2017, Mastrogiuseppe2017} it was possible to further simplify the self-consistent equations because the resulting analogous equation to Eqs~\eqref{2syn} and Eq~\eqref{2adap} was a conservative system. However, in the networks studied here, synaptic filtering and adaptation add the convolutional terms in Eqs~\eqref{2syn} and Eq~\eqref{2adap} that make the system non-conservative. Therefore, we followed an alternative approach and found the solutions to the self-consistent equations using an iterative scheme, that circumvents solving directly the integral equations.

\subparagraph{Iterative scheme}
We solve the self-consistent equations numerically following a single-unit iterative scheme, as in \cite{Lerchner2006a, Dummer2014, WieBer15, Stern2014}:
\begin{itemize}
\item First, we simulate the dynamics in Eqs~\eqref{DMFT_syn} and~\eqref{DMFT_adap} assuming white Gaussian noise with a certain mean $\left[\xi\right]^{\left(0\right)}$ and autocorrelation function $\left[\xi\left(t\right)\xi\left(t+\tau\right)\right] = \left(\sigma_\xi^{\left(0\right)}\right)^2 \delta\left(\tau\right)$.

\item We calculate the autocorrelation functions of the firing rate and input currents empirically, $\mu^{\left(0\right)}$, $\Delta^{\left(0\right)}$, $\left\langle \phi \right\rangle^{\left(0\right)}$ and $C^{\left(0\right)}\left(\tau\right)$.

\item We simulate in the new iteration $k+1$ the noise following the self-consistent statistics obtained in the previous iteration, as indicated by Eqs~\eqref{selfcon1} and \eqref{selfcon2}

\begin{eqnarray}\label{iter_method1}
\left[\xi\right]^{\left(k+1\right)} &=& J\left(C_E-gC_I\right) \left\langle \phi\right\rangle^{\left(k\right)}\\
\left[\xi\left(t\right)\xi\left(t+\tau\right)\right]^{k+1} &=& J^2\left(C_E+g^2C_I\right) \left(C^{\left(k\right)} \left(\tau\right) - \left\langle \phi\right\rangle^{\left(k\right)}\right). \label{iter_method11}
\end{eqnarray}
 
 \noindent In order to numerically generate a Gaussian variable with autocorrelation function $G\left(\tau\right)$, we first generate the noise in the Fourier domain, where each frequency component of the noise is given by
 
 \begin{equation}
 \tilde{\xi}\left(\omega\right) = \sqrt{\tilde{G}\left(\omega\right)} e^{i\psi},
 \end{equation}
 \noindent where $\tilde{G}\left(\omega\right)$ denotes the Fourier transform of the target autocorrelation function, and $\psi$ is a random variable with uniform probability density in the range $\left[-\pi, \pi\right]$.

\item We repeat the previous step until the values $\mu^{\left(k\right)}$, $\Delta^{\left(k\right)}$, $\left\langle \phi \right\rangle^{\left(k\right)}$ and $C^{\left(k\right)}\left(\tau\right)$ do not vary beyond a certain tolerance for new iterations.
\end{itemize}

We find that such an iterative method applied to the systems studied here converges to a solution for the parameters of the noise after a few iterations, independently of the noise properties used in the initial step. 

The drawbacks of this iterative scheme are that the two-dimensional system needs to be simulated several times at each iteration in order to determine the first and second order statistics of the input current and the firing rate, which is in general a computationally costly operation. We also find that the method converges more robustly to the solution (given the fact that both the trial length in the simulation and the number of trials are finite), at the expense of initial speed convergence, when the first and second moments of the noise are only partially updated at each iteration,  so that

\begin{eqnarray}
\left[\xi\right]^{\left(k+1\right)} &=& (1-\alpha)\left[\xi\right]^{\left(k\right)} + \alpha J\left(C_E-gC_I\right) \left\langle \phi\right\rangle^{\left(k\right)}, \label{iter_method2}
\end{eqnarray}

\noindent and similarly for the second-moment equation, where $\alpha$ is a parameter between zero and one. In this work, we used $\alpha=0.6$.

This method is inefficient for very large adaptation and synaptic time constants, since it requires simulating with both a fine temporal resolution (faster than the membrane time constant) over very large intervals (much larger than the slow adaptive/synaptic timescale). Another drawback of the iterative method is that its convergence is based on the assumption that smooth changes in the noise statistics lead to smooth changes in the statistics of the firing rates. In general, close to a bifurcation, this requirement may not hold. 

 \subparagraph{Dynamics with intrinsic noise}
 
We next study how white Gaussian noise, independent between neurons and intrinsic to each unit in the network, affects the dynamics of the system. On the mean-field level, this is equivalent to studying a network where each neuron spikes at a Poisson process whose rate varies in time as $\phi\left(x_i\left(t\right)\right)$ \cite{Kadmon2015}. The additional input to each neuron, whose dynamics are given in Eqs~\eqref{sys_adap1} and \eqref{sys_adap2}, is now  

\begin{equation}
I^{ext}_i\left(t\right) = \eta_i\left(t\right),
\end{equation}

where $\left[\eta_i\right]=0$, and $\left[\eta_i\left(t\right)\eta_j\left(t+\tau\right)\right]=\delta_{ij} \frac{\sigma_\eta^2}{2} \delta\left(\tau \right)$, and Gaussian distributed. The DMF equations are derived following the same steps as in the absence of intrinsic noise. The stochastic variable $\xi\left(t\right)$ is the sum of the recurrent input and the intrinsic noise. Its first moment remains unchanged:

\begin{align}
\left[\xi\left(t\right)\right] &= \left\langle\sum_{j=1}^N J_{ij} \phi\left(x_j \left(t\right)\right) + \eta_i\left(t\right) \right\rangle\\
&= J\left(C_E-gC_I\right) \left\langle \phi \right\rangle,
\end{align}

\noindent which is the same result as Eq~\eqref{selfcon1}. The second moment of the stochastic process is the sum of the variance generated by the recurrent connections and the variance of the intrinsic noise

\begin{align}
\left[\xi\left(t\right)\xi\left(t+\tau\right)\right] &=  \left\langle \sum_{k=1}^N J_{ik} \phi_k \left(t\right) \sum_{l=1}^N  J_{il} \phi_l \left(t\right) + \eta_i\left(t\right) \eta_i\left(t+\tau\right)\right\rangle\\
& = J^2\left(C_E+g^2C_I\right) \left(C\left(\tau\right) - \left\langle \phi\right\rangle^2\right) + \frac{1}{2}\sigma_\eta^2 \delta\left(\tau\right).
\end{align}

Accordingly, the iterative scheme now takes into account the equation above, so that the equation for the second moment of the self-consistent relation (Eq~\ref{iter_method11}) reads when there is intrinsic noise

\begin{equation}\label{iter_methodnoise}
\left[\xi\left(t\right)\xi\left(t+\tau\right)\right]^{\left(k+1\right)} = J^2\left(C_E+g^2C_I\right) \left(C^{\left(k\right)} \left(\tau\right) - \left\langle \phi\right\rangle^{\left(k\right)}\right) + \frac{1}{2}\sigma_\eta^2 \delta\left(\tau\right). 
\end{equation}

Adding white noise produces a discontinuity in the derivative of the autocorrelation function of the firing rates at zero lag (Fig~\ref{fig7} A and D). This can be shown by integrating explicitly both sides of the Eqs \eqref{2syn} and~\eqref{2adap} around zero when the external noise is added. It results in the condition

\begin{equation}
\dot{\Delta} \left(0^+\right) - \dot{\Delta} \left(0^-\right) = \frac{1}{2}\sigma_\eta^2.
\end{equation} 

Since the autocorrelation function is a symmetric function, $\dot{\Delta}\left(0^+\right) = -\dot{\Delta}\left(0^+\right)$, leading to

\begin{equation}
\dot{ \Delta}_0 = \sigma_\eta^2.
\end{equation}

\noindent Thus, the autocorrelation function of the input current decays linearly at zero time lag with a slope proportional to the external noise intensity, which also extends to the autocorrelation function of the firing rate. 

\subsection*{Definition of the timescale of the activity}

The activity of multivariable dynamical systems ranges over several timescales. In particular, for stable linear systems, the timescales of the activity are given by the inverse of the absolute values of the real part of the eigenvalues. As we showed before, for single adaptive or synaptic neurons, the activity consists of two modes that evolve at two different timescales. However, the relative contribution of each of the excited modes can make one timescale more predominant than the other, as it happens for slow adaptation time constant, which becomes effectively undetectable in the single neuron dynamics.

In this work, we calculate the timescale of the activity for linear systems as the average of the timescales of the activated input current modes, weighed by their contribution (Fig~\ref{fig1}). For a linear system with filter $h\left(t\right) = \sum_k a_k e^{-\frac{t}{\tau_k}}$, the correlation time is

\begin{equation}
\tau_{corr} = \frac{\sum_k \left\lvert a_k\right\rvert \tau_k}{\sum_k \left\lvert a_k\right\rvert}.
\end{equation}

For large networks, which are high-dimensional non-linear systems, we define the main timescale of the activity as the time lag at which the autocorrelation function has decayed to a fraction $e^{-\frac{1}{2}}$ of its maximum (Figs~\ref{fig5} and~\ref{fig7}):

\begin{equation}
\tau_{corr} = 2 \cdot \underset{\tau}{\operatorname{argmin}}  \left\lvert E\left[C\left(\tau\right)\right] - \frac{E\left[C\left(\tau\right)\right]}{\sqrt{e}} \right\rvert,
\end{equation}

where $E[C\left(\tau\right)]$ is the envelope of the autocorrelation function, calculated as the norm of its analytic signal, computed using the Hilbert transform. This corresponds to the width of the envelope at which the autocorrelation decays to $e^{-0.5}$ of its value. For an exponentially decaying correlation function, this measure corresponds to the decay time constant. For a Gaussian envelope, this measure would correspond to two times its standard deviation, $2 \sigma$.


\section*{Acknowledgments}
This work was funded by the Ecole des Neurosciences de Paris Ile-de-France, the Programme Emergences of City of Paris, Agence Nationale de la Recherche grants
ANR-16-CE37-0016-01 and ANR-17-ERC2-0005-01, and the program “Investissements d’Avenir” launched by the
French Government and implemented by the ANR, with the references ANR-10-LABX-0087 IEC and ANR11-IDEX-0001-02
PSL* Research University. The funders had no role in study design, data collection and
analysis, decision to publish, or preparation of the manuscript.


\begin{thebibliography}{10}
	
	\bibitem{Fairhall2001}
	Fairhall AL, Lewen GD, Bialek W, {De Ruyter van Steveninck} RR.
	\newblock {Efficiency and ambiguity in an adaptive neural code}.
	\newblock Nature. 2001;412(6849):787--792.
	\newblock doi:{10.1038/35090500}.
	
	\bibitem{Grothe2010}
	Grothe B, Pecka M, McAlpine D.
	\newblock {Mechanisms of Sound Localization in Mammals}.
	\newblock Physiological Reviews. 2010;90(3):983--1012.
	\newblock doi:{10.1152/physrev.00026.2009}.
	
	\bibitem{Tchumatchenko}
	Tchumatchenko T, Malyshev A, Wolf F, Volgushev M.
	\newblock {Ultrafast Population Encoding by Cortical Neurons}.
	\newblock Journal of Neuroscience. 2011;31(34):12171--12179.
	\newblock doi:{10.1523/JNEUROSCI.2182-11.2011}.
	
\bibitem{Smith2004}
Smith PL, Ratcliff R. 
\newblock {Psychology and neurobiology of simple decisions}.
\newblock Trends in Neurosciences. 2004;27(3):161--168
\newblock doi:{10.1016/j.tins.2004.01.006}.
	
	\bibitem{Miyashita1988}
	Miyashita Y, Chang HS.
	\newblock {Neuronal correlate of pictorial short-term memory in the primate
		temporal cortex}.
	\newblock Nature. 1988;331(6151):68--70.
	\newblock doi:{10.1038/331068a0}.
	
	\bibitem{Bair2004}
	Bair W, Movshon JA.
	\newblock {Adaptive Temporal Integration of Motion in Direction-Selective
		Neurons in Macaque Visual Cortex}.
	\newblock Journal of Neuroscience. 2004;24(33):7305--7323.
	\newblock doi:{10.1523/JNEUROSCI.0554-04.2004}.
	
	\bibitem{Bernacchia2011}
	Bernacchia A, Seo H, Lee D, Wang XJ.
	\newblock {A reservoir of time constants for memory traces in cortical
		neurons}.
	\newblock Nature Neuroscience. 2011;14(3):366--372.
	\newblock doi:{10.1038/nn.2752}.
	
	\bibitem{Murray2014}
	Murray JD, Bernacchia A, Freedman DJ, Romo R, Wallis JD, Cai X, et~al.
	\newblock {A hierarchy of intrinsic timescales across primate cortex}.
	\newblock Nature Neuroscience. 2014;17(12):1661--1663.
	\newblock doi:{10.1038/nn.3862}.
	
\bibitem{Wang2001}
Wang XJ. 
\newblock {Synaptic reverberation underlying mnemonic persistent activity};
\newblock Trends in Neurosciences. 2001;24(8):455--643.
\newblock doi:{10.1016/S0166-2236(00)01868-3}.

\bibitem{Wang2008}
Wang XJ. 
\newblock {Decision Making in Recurrent Neuronal Circuits}; 
\newblock Neuron. 2008;60(2):215--234.
\newblock doi:{10.1016/j.neuron.2008.09.034}.

	\bibitem{Litwin-Kumar2012}
	Litwin-Kumar A, Doiron B.
	\newblock {Slow dynamics and high variability in balanced cortical networks
		with clustered connections}.
	\newblock Nature Neuroscience. 2012;15(11):1498--1505.
	\newblock doi:{10.1038/nn.3220}.
	
	\bibitem{Huang2017}
	Huang C, Doiron B.
	\newblock {Once upon a (slow) time in the land of recurrent neuronal
		networks{\ldots}}
	\newblock Current Opinion in Neurobiology. 2017;46:31--38.
	\newblock doi:{10.1016/J.CONB.2017.07.003}.
	
	\bibitem{Buonomano2009}
	Buonomano DV, Maass W.
	\newblock {State-dependent computations: Spatiotemporal processing in cortical
		networks}.
	\newblock Nature Reviews Neuroscience. 2009;10(2):113--125.
	\newblock doi:{10.1038/nrn2558}.
	
	\bibitem{Zucker2002}
	Zucker RS, Regehr WG.
	\newblock {Short-Term Synaptic Plasticity}.
	\newblock Annual Review of Physiology. 2002;64(1):355--405.
	\newblock doi:{10.1146/annurev.physiol.64.092501.114547}.
	
	\bibitem{Markram1998}
	Markram H, Wang Y, Tsodyks M.
	\newblock {Differential signaling via the same axon of neocortical pyramidal
		neurons}.
	\newblock Proceedings of the National Academy of Sciences.
	1998;doi:{10.1073/pnas.95.9.5323}.
	
	\bibitem{Newberry}
	Newberry NR, Nicoll RA.
	\newblock {Direct hyperpolarizing action of baclofen on hippocampal pyramidal
		cells.}
	\newblock Nature;308(5958):450--2.
	\newblock doi:{10.1038/308450a0}
	
	\bibitem{Batchelor1994}
	Batchelor AM, Madge DJ, Garthwaite J.
	\newblock {Synaptic activation of metabotropic glutamate receptors in the
		parallel Fibre-Purkinje cell pathway in rat cerebellar slices}.
	\newblock Neuroscience. 1994;63(4):911--915.
	\newblock doi:{10.1016/0306-4522(94)90558-4}.
	
	\bibitem{Garthwaite1991}
	Garthwaite J.
	\newblock {Glutamate, nitric oxide and cell-cell signalling in the nervous
		system.}
	\newblock Trends in neurosciences. 1991;14(2):60--7.
	\newblock doi:{10.1016/0166-2236(91)90022-M}
	
	\bibitem{Lester1990}
	Lester RAJ, Clements JD, Westbrook GL, Jahr CE.
	\newblock {Channel kinetics determine the time course of NMDA receptor-mediated
		synaptic currents}.
	\newblock Nature. 1990;346(6284):565--567.
	\newblock doi:{10.1038/346565a0}.
	
	\bibitem{Johnston1995}
	Johnston D, Wu SMs.
	\newblock {Foundations of cellular neurophysiology}.
	\newblock MIT Press; 1995.
	
	\bibitem{Berridge2003}
	Berridge MJ, Bootman MD, Roderick HL. {Calcium: Calcium signalling: Dynamics,
		homeostasis and remodelling}; 2003.
	\newblock Nature Reviews Molecular cell biology. 2003;4(7):517.
	\newblock doi:{10.1038/nrm1155}.
	
	\bibitem{Gal2010}
	Gal A, Eytan D, Wallach A, Sandler M, Schiller J, Marom S.
	\newblock {Dynamics of Excitability over Extended Timescales in Cultured
		Cortical Neurons}.
	\newblock Journal of Neuroscience. 2010;30(48):16332--16342.
	\newblock doi:{10.1523/JNEUROSCI.4859-10.2010}.
	
	\bibitem{LaCamera2006}
	{La Camera} G, Rauch A, Thurbon D, L{\"{u}}scher HR, Senn W, Fusi S.
	\newblock {Multiple Time Scales of Temporal Response in Pyramidal and Fast
		Spiking Cortical Neurons}.
	\newblock Journal of Neurophysiology. 2006;96(6):3448--3464.
	\newblock doi:{10.1152/jn.00453.2006}.
	
	\bibitem{Benda2003}
	Benda J, Herz AVM.
	\newblock {A Universal Model for Spike-Frequency Adaptation}.
	\newblock Neural Computation. 2003;15(11):2523--2564.
	\newblock doi:{10.1162/089976603322385063}.
	
	\bibitem{Ermentrout2001}
	Ermentrout B, Pascal M, Gutkin B.
	\newblock {The effects of spike frequency adaptation and negative feedback on
		the synchronization of neural oscillators}.
	\newblock Neural Computation. 2001;13(6):1285--1310.
	\newblock doi:{10.1162/08997660152002861}.
	
	\bibitem{Hennig2013}
	Hennig MH.
	\newblock {Theoretical models of synaptic short term plasticity}.
	\newblock Frontiers in Computational Neuroscience. 2013;7.
	\newblock doi:{10.3389/fncom.2013.00154}.
	
	\bibitem{Wilson1972}
	Wilson HR, Cowan JD.
	\newblock {Excitatory and Inhibitory Interactions in Localized Populations of
		Model Neurons}.
	\newblock Biophysical Journal. 1972;12(1):1--24.
	\newblock doi:{10.1016/S0006-3495(72)86068-5}.
	
	\bibitem{Wilson1973}
	Wilson HR, Cowan JD.
	\newblock {A mathematical theory of the functional dynamics of cortical and
		thalamic nervous tissue}.
	\newblock Kybernetik. 1973;13(2):55--80.
	\newblock doi:{10.1007/BF00288786}.
	
	\bibitem{Sompolinsky1988}
	Sompolinsky H, Crisanti A, Sommers HJ.
	\newblock {Chaos in random neural networks}.
	\newblock Physical Review Letters. 1988;61(3):259--262.
	\newblock doi:{10.1103/PhysRevLett.61.259}.
	
	\bibitem{Abbott1994}
	Abbott LF.
	\newblock {Decoding neuronal firing and modelling neural networks}.
	\newblock Quarterly Reviews of Biophysics. 1994;27(3):291--331.
	\newblock doi:{10.1017/S0033583500003024}.
	
	\bibitem{Amit1997}
	Amit D, Brunel N.
	\newblock {Model of global spontaneous activity and local structured activity
		during delay periods in the cerebral cortex}.
	\newblock Cerebral Cortex. 1997;7(3):237--252.
	\newblock doi:{10.1093/cercor/7.3.237}.
	
	\bibitem{Brunel2000}
	Brunel N.
	\newblock {Dynamics of networks of randomly connected excitatory and inhibitory
		spiking neurons}.
	\newblock Journal of Physiology Paris. 2000;94(5-6):445--463.
	\newblock doi:{10.1016/S0928-4257(00)01084-6}.
	
	\bibitem{Mastrogiuseppe2017}
	Mastrogiuseppe F, Ostojic S.
	\newblock {Intrinsically-generated fluctuating activity in
		excitatory-inhibitory networks}. 
	\newblock PLoS Computational Biology. 2017;13(4):1--40.
	\newblock doi:{10.1371/journal.pcbi.1005498}.
	
	
	\bibitem{Rajan2010}
	Rajan K, Abbott LF, Sompolinsky H.
	\newblock {Stimulus-dependent suppression of chaos in recurrent neural
		networks}.
	\newblock Physical Review E.
	2010;82(1):1--5.
	\newblock doi:{10.1103/PhysRevE.82.011903}.
	
	\bibitem{Kadmon2015}
	Kadmon J, Sompolinsky H.
	\newblock {Transition to chaos in random neuronal networks}.
	\newblock Physical Review X. 2015;5(4).
	\newblock doi:{10.1103/PhysRevX.5.041030}.
	
	\bibitem{Rajan2006}
	Rajan K, Abbott LF.
	\newblock {Eigenvalue spectra of random matrices for neural networks}.
	\newblock Physical Review Letters. 2006;97(18):2--5.
	\newblock doi:{10.1103/PhysRevLett.97.188104}.
	
	\bibitem{Bimbard2016}
	Bimbard C, Ledoux E, Ostojic S.
	\newblock {Instability to a heterogeneous oscillatory state in randomly
		connected recurrent networks with delayed interactions}.
	\newblock Physical Review E. 2016;94(6):3--8.
	\newblock doi:{10.1103/PhysRevE.94.062207}.
	
	\bibitem{Schuecker2017}
	Schuecker J, Goedeke S, Helias M.
	\newblock {Optimal sequence memory in driven random networks};
	\newblock Physical Review X. 2018;8(4):041029.
	\newblock doi:{10.1103/PhysRevX.8.041029}
	
	
	\bibitem{Aljadeff2015}
	Aljadeff J, Stern M, Sharpee T.
	\newblock {Transition to Chaos in Random Networks with Cell-Type-Specific
		Connectivity};
	\newblock Physical Review Letters. 2015;114(8).
	\newblock doi:{10.1103/PhysRevLett.114.088101}.
	
	\bibitem{Harish2015}
	Harish O, Hansel D.
	\newblock {Asynchronous Rate Chaos in Spiking Neuronal Circuits}.
	\newblock PLoS Computational Biology. 2015;11(7):e1004266.
	\newblock doi:{10.1371/journal.pcbi.1004266}.
	
	\bibitem{Stern2014}
	Stern M, Sompolinsky H, Abbott LF.
	\newblock {Dynamics of random neural networks with bistable units}.
	\newblock Physical Review E.
	2014;90(6):1--7.
	\newblock doi:{10.1103/PhysRevE.90.062710}.
	
	\bibitem{Lerchner2006a}
	Lerchner A, Sterner G, Hertz J, Ahmadi M.
	\newblock {Mean field theory for a balanced hypercolumn model of orientation
		selectivity in primary visual cortex}.
	\newblock Network: Computation in Neural Systems. 2006;17(2):131--150.
	\newblock doi:{10.1080/09548980500444933}.
	
	\bibitem{Dummer2014}
	Dummer B, Wieland S, Lindner B.
	\newblock {Self-consistent determination of the spike-train power spectrum in a
		neural network with sparse connectivity}.
	\newblock Frontiers in Computational Neuroscience. 2014;8(September):1--12.
	\newblock doi:{10.3389/fncom.2014.00104}.
	
	\bibitem{WieBer15}
	Wieland S, Bernardi D, Schwalger T, Lindner B.
	\newblock {Slow fluctuations in recurrent networks of spiking neurons}.
	\newblock Phys Rev E. 2015;92:040901(R).
	\newblock doi:{10.1103/PhysRevE.92.040901}
	
	\bibitem{Ostojic2011}
	Ostojic S, Brunel N.
	\newblock {From Spiking Neuron Models to Linear-Nonlinear Models}.
	\newblock PLoS Computational Biology. 2011;7(1):e1001056.
	\newblock doi:{10.1371/journal.pcbi.1001056}.
	
	\bibitem{Troyer1997}
	Troyer TW, Miller KD.
	\newblock {Physiological Gain Leads to High ISI Variability in a Simple Model
		of a Cortical Regular Spiking Cell}.
	\newblock Neural Computation. 1997;9(5):971--983.
	\newblock doi:{10.1162/neco.1997.9.5.971}.
	
	\bibitem{Murphy2009}
	Murphy BK, Miller KD.
	\newblock {Balanced amplification: a new mechanism of selective amplification
		of neural activity patterns.}
	\newblock Neuron. 2009;61(4):635--48.
	\newblock doi:{10.1016/j.neuron.2009.02.005}.
	
	\bibitem{Ahmadian2013}
	Ahmadian Y, Rubin DB, Miller KD.
	\newblock {Analysis of the stabilized supralinear network.}
	\newblock Neural computation. 2013;25(8):1994--2037.
	\newblock doi:{10.1162/NECO\_a\_00472}.
	
	\bibitem{Jaeger2001}
	Jaeger H.
	\newblock {The "echo state" approach to analysing and training recurrent neural
		networks}.
	\newblock In: GMD-Forschungszentrum Informationstechnik Report 148; 2001.
	
	\bibitem{Sussillo2009}
	Sussillo D, Abbott LF.
	\newblock {Generating Coherent Patterns of Activity from Chaotic Neural
		Networks}.
	\newblock Neuron. 2009;63(4):544--557.
	\newblock doi:{10.1016/j.neuron.2009.07.018}.
	
	\bibitem{Brown2000}
	Brown DA.
	\newblock {M-Current: From Discovery to Single Channel Currents}.
	\newblock In: Slow Synaptic Responses and Modulation. Tokyo: Springer Japan;
	2000. p. 15--26.
	
	\bibitem{Stanley2011}
	Stanley DA, Bardakjian BL, Spano ML, Ditto WL.
	\newblock {Stochastic amplification of calcium-activated potassium currents in
		Ca2+ microdomains}.
	\newblock Journal of Computational Neuroscience. 2011;31(3):647--666.
	\newblock doi:{10.1007/s10827-011-0328-x}.
	
	\bibitem{Naud2008}
	Naud R, Marcille N, Clopath C, Gerstner W.
	\newblock {Firing patterns in the adaptive exponential integrate-and-fire
		model}.
	\newblock Biological Cybernetics. 2008;99(4-5):335--347.
	\newblock doi:{10.1007/s00422-008-0264-7}.
	
	\bibitem{Schwalger2010}
	Schwalger T, Fisch K, Benda J, Lindner B.
	\newblock {How noisy adaptation of neurons shapes interspike interval
		histograms and correlations}.
	\newblock PLoS Computational Biology. 2010;6(12):e1001026.
	\newblock doi:{10.1371/journal.pcbi.1001026}.
	
	\bibitem{Ladenbauer2014}
	Ladenbauer J, Augustin M, Obermayer K.
	\newblock {How adaptation currents change threshold, gain and variability of
		neuronal spiking}.
	\newblock Journal of Neurophysiology. 2013;111(5):939--953.
	\newblock doi:{10.1152/jn.00586.2013}.
	
	\bibitem{Richardson2003}
	Richardson MJE, Brunel N, Hakim V.
	\newblock {From Subthreshold to Firing-Rate Resonance}.
	\newblock Journal of Neurophysiology. 2003;89(5):2538--2554.
	\newblock doi:{10.1152/jn.00955.2002}.
	
	\bibitem{Brunel2003}
	Brunel N, Hakim V, Richardson MJE.
	\newblock {Firing-rate resonance in a generalized integrate-and-fire neuron
		with subthreshold resonance}.
	\newblock Physical Review E. 2003;67(5):051916.
	\newblock doi:{10.1103/PhysRevE.67.051916}.
	
	\bibitem{Ladenbauer2012}
	Ladenbauer J, Augustin M, Shiau LJ, Obermayer K.
	\newblock {Impact of adaptation currents on synchronization of coupled
		exponential integrate-and-fire neurons}.
	\newblock PLoS Computational Biology. 2012;8(4).
	\newblock doi:{10.1371/journal.pcbi.1002478}.
	
	\bibitem{Augustin2013}
	Augustin M, Ladenbauer J, Obermayer K.
	\newblock {How adaptation shapes spike rate oscillations in recurrent neuronal
		networks}.
	\newblock Frontiers in Computational Neuroscience. 2013;7:9.
	\newblock doi:{10.3389/fncom.2013.00009}.
	
	\bibitem{Schwalger2013}
	Schwalger T, Lindner B.
	\newblock {Patterns of interval correlations in neural oscillators with
		adaptation}.
	\newblock Frontiers in Computational Neuroscience. 2013;7:164.
	\newblock doi:{10.3389/fncom.2013.00164}.
	
	\bibitem{Laing2002}
	Laing CR, Chow CC.
	\newblock {A Spiking Neuron Model for Binocular Rivalry}.
	\newblock Journal of Computational Neuroscience. 2002;12(1):39--53.
	\newblock doi:{10.1023/A:1014942129705}.
	
	\bibitem{Naud2012}
	Naud R, Gerstner W.
	\newblock {Coding and Decoding with Adapting Neurons: A Population Approach to
		the Peri-Stimulus Time Histogram}.
	\newblock PLoS Computational Biology. 2012;8(10):e1002711.
	\newblock doi:{10.1371/journal.pcbi.1002711}.
	
	\bibitem{Pozzorini2013}
	Pozzorini C, Naud R, Mensi S, Gerstner W.
	\newblock {Temporal whitening by power-law adaptation in neocortical neurons}.
	\newblock Nature Neuroscience. 2013;16(7):942--948.
	\newblock doi:{10.1038/nn.3431}.
	
	\bibitem{augustin2017}
	Augustin M, Ladenbauer J, Baumann F, Obermayer K.
	\newblock {Low-dimensional spike rate models derived from networks of adaptive
		integrate-and-fire neurons: Comparison and implementation}.
	\newblock PLoS Computational Biology. 2017;13(6):1--46.
	\newblock doi:{10.1371/journal.pcbi.1005545}.
	
	\bibitem{Sjostrom2001}
	Sj{\"{o}}str{\"{o}}m PJ, Turrigiano GG, Nelson SB.
	\newblock {Rate, Timing, and Cooperativity Jointly Determine Cortical Synaptic
		Plasticity}.
	\newblock Neuron. 2001;32(6):1149--1164.
	\newblock doi:{10.1016/S0896-6273(01)00542-6}.
	
	\bibitem{Ko2011}
	Ko H, Hofer SB, Pichler B, Buchanan KA, Sj{\"{o}}str{\"{o}}m PJ, Mrsic-Flogel
	TD.
	\newblock {Functional specificity of local synaptic connections in neocortical
		networks}.
	\newblock Nature. 2011;473(7345):87--91.
	\newblock doi:{10.1038/nature09880}.
	
	\bibitem{Marti2018}
	Mart{\'{i}} D, Brunel N, Ostojic S.
	\newblock {Correlations between synapses in pairs of neurons slow down dynamics
		in randomly connected neural networks}.
	\newblock Physical Review E. 2018;97(6):062314.
	\newblock doi:{10.1103/PhysRevE.97.062314}.
	
	\bibitem{Mastrogiuseppe2018}
	Mastrogiuseppe F, Ostojic S.
	\newblock {Linking Connectivity, Dynamics, and Computations in Low-Rank
		Recurrent Neural Networks.}
	\newblock Neuron. 2018;99(3):609--623.e29.
	\newblock doi:{10.1016/j.neuron.2018.07.003}.
	
	\bibitem{Sussillo2014}
	Sussillo D.
	\newblock {Neural circuits as computational dynamical systems}.
	\newblock Current Opinion in Neurobiology. 2014;25:156--163.
	\newblock doi:{10.1016/j.conb.2014.01.008}.
	
	\bibitem{Barak2017}
	Barak O.
	\newblock {Recurrent neural networks as versatile tools of neuroscience
		research}.
	\newblock Current Opinion in Neurobiology. 2017;46:1--6.
	\newblock doi:{10.1016/J.CONB.2017.06.003}.
	
	\bibitem{Nicola2016}
	Nicola W, Clopath C.
	\newblock {Supervised learning in spiking neural networks with FORCE training}.
	\newblock Nature Communications. 2017;8(1):2208.
	\newblock doi:{10.1038/s41467-017-01827-3}.
	
	\bibitem{Bellec2018}
	Bellec G, Salaj D, Subramoney A, Legenstein R, Maass W.
	\newblock {Long short-term memory and learning-to-learn in networks of spiking
		neurons}. 
	\newblock arXiv preprint. 2018;arXiv:1803.09574.
	
	
	\bibitem{Ostojic2014}
	Ostojic S.
	\newblock {Two types of asynchronous activity in networks of excitatory and
		inhibitory spiking neurons}.
	\newblock Nature Neuroscience. 2014;17(4):594--600.
	\newblock doi:{10.1038/nn.3658}.
	
\end{thebibliography}
\end{document}